\documentclass[aps,pre,10pt,showpacs,floatfix,onecolumn,nofootinbib,a4paper]{revtex4-2}

\usepackage{amssymb, amsmath, amsthm, mathtools}
\usepackage{bm}
\usepackage{mathrsfs}
\usepackage{hyperref,bookmark}
\usepackage{graphicx}
\usepackage{epsfig,color}
\usepackage{float}
\usepackage{subcaption}
\usepackage{verbatim}
\usepackage{dcolumn}
\usepackage{tikz}
\usepackage{url}
\usepackage{cases}
\usepackage[utf8x]{inputenc}
\usepackage[english]{babel}
\usepackage{epsfig}
\usepackage{listings}
\usepackage{paralist}
\usepackage{mdwlist}
\usepackage{bm}
\usepackage{dsfont}
\usepackage{oldgerm}
\usepackage{geometry}
\usepackage{relsize}
\usepackage{afterpage}
\usepackage{lmodern}
\usepackage{tabularx}
\usepackage{cancel}
\usepackage{dutchcal} 
\usepackage{natbib} 

\newcommand{\dd}{\mathrm{d}}

\newcommand{\vt}{\vartheta}
\newcommand{\vp}{\varphi}
\newcommand{\trans}{^\mathsf{T}}
\newcommand{\transl}{\mathscr{V}}
\newcommand{\av}{\bm{a}}
\newcommand{\n}{\bm{n}}
\newcommand{\cv}{\bm{c}}

\newcommand{\dv}{\bm{d}}
\newcommand{\e}{\bm{e}}
\newcommand{\x}{\bm{x}}
\newcommand{\zero}{\bm{0}}
\newcommand{\y}{\bm{y}}

\newcommand{\normal}{\bm{\nu}}
\newcommand{\surface}{\mathscr{S}}
\newcommand{\tplane}{\mathscr{T}}

\newcommand{\nablas}{\nabla\!_\mathrm{s}}
\newcommand{\nablac}{\nabla\!_\mathrm{c}}
\newcommand{\nablast}{\nabla^\ast\!\!\!_\mathrm{s}}

\newcommand{\nablastwo}{\nabla^2\!\!\!\!_\mathrm{s}\,}

\newcommand{\uv}{\bm{u}}

\newcommand{\h}{\bm{h}}
\newcommand{\rv}{\bm{r}}

\newcommand{\vv}{\bm{v}}

\newcommand{\W}{\mathbf{W}}

\renewcommand{\H}{\mathbf{H}}
\newcommand{\I}{\mathbf{I}}
\newcommand{\D}{\mathbf{D}}
\newcommand{\R}{\mathbf{R}}

\newcommand{\U}{\mathbf{U}}

\newcommand{\curl}{\operatorname{curl}}
\newcommand{\curls}{\curl_\mathrm{s}\!}
\newcommand{\curlc}{\curl_\mathrm{c}\!}
\newcommand{\divs}{\diver_\mathrm{s}\!}
\newcommand{\divc}{\diver_\mathrm{c}\!}
\newcommand{\tr}{\operatorname{tr}}

\newcommand{\skw}{\operatorname{skw}}

\newcommand{\curvature}{(\nablas\normal)}

\newcommand{\euclid}{\mathscr{E}}
\newcommand{\framec}{(\e_1,\e_2,\e_3)}
\newcommand{\frameen}{(\e_1,\e_2,\normal)}
\newcommand{\framed}{(\n_1,\n_2,\n)}

\newcommand{\framen}{(\n,\n_\perp,\normal)}
\newcommand{\framenast}{(\n^\ast,\n_\perp^\ast,\normal^\ast)}
\newcommand{\framee}{(\e_u,\e_v,\normal)}

\newcommand{\framet}{(\tangent,\tangent_\perp,\normal)}
\newcommand{\tangent}{\bm{t}}
\newcommand{\tperp}{\bm{t}_\perp}
\newcommand{\nperp}{\n_\perp}

\newcommand{\curve}{\bm{x}}
\newcommand{\Curve}{\mathscr{C}}

\newcommand{\orth}{\mathsf{SO}(3)}

\newcommand{\proj}{\mathbf{P}(\normal)}

\newcommand{\Proj}{\mathbf{P}}

\newcommand{\diver}{\operatorname{div}} 

\newcommand{\kn}{\kappa_\mathrm{n}}
\newcommand{\kg}{\kappa_\mathrm{g}}
\newcommand{\tg}{\tau_\mathrm{g}}
\newcommand{\tgp}{\tau_\mathrm{g}^\perp}
\newcommand{\kgp}{\kappa_\mathrm{g}^\perp}
\newcommand{\knp}{\kappa_\mathrm{n}^\perp}
\newcommand{\rvu}{\bm{r}_u}
\newcommand{\rvv}{\bm{r}_v}
\newcommand{\spin}{\bm{\Omega}}
\newcommand{\bend}{\bm{b}}

\theoremstyle{definition}

\begin{document}
	\latintext
	\title{Surface  Nematic Uniformity}
	\author{Andrea Pedrini}
	\email{andrea.pedrini@unipv.it}
	\affiliation{Dipartimento di Matematica, Universit\`a di Pavia, Via Ferrata 5, 27100 Pavia, Italy}
	\author{Epifanio G. Virga}
	\email{eg.virga@unipv.it}
	\affiliation{Dipartimento di Matematica, Universit\`a di Pavia, Via Ferrata 5, 27100 Pavia, Italy}

	\date{\today}

	\begin{abstract}
An ant-like observer confined to a two-dimensional surface traversed by stripes would wonder whether this striped landscape could be devised in such a way as to appear to be the same wherever they go. Differently stated, this is the problem studied in this paper. In a more technical jargon, we determine all possible \emph{uniform} nematic fields on a smooth surface. It was already known that for such a field to exist, the surface must have constant \emph{negative} Gaussian curvature. Here, we show that all uniform nematic fields on such a surface are parallel transported (in Levi-Civita's sense) by special systems of geodesics, which (with scant inventiveness) are termed \emph{uniform}. We prove that, for every geodesic on the surface, there are \emph{two} systems of uniform geodesics that include it; they are conventionally called \emph{right} and \emph{left}, to allude at a possible intrinsic definition of handedness. We found explicitly all uniform fields for Beltrami's pseudosphere. Since both geodesics and uniformity are preserved under isometries, by a classical theorem of Minding, the solution for the pseudosphere carries over all other admissible surfaces, thus providing a general solution to the problem (at least in principle).
	\end{abstract}

	\maketitle

\section{Introduction}\label{sec:intro}
Owing to its etymology, \emph{nematic} refers to anything related to threads. A line field on a curved surface is thus an example of nematic system as good as a liquid crystal shell, where elongated molecules condensed in an ordered phase cover a curved substrate.
This paper is concerned with a general property of unit vector fields $\n$ tangent to a surface $\surface$, a property that is predominantly geometric, but with a physical meaning.

It is the condition of \emph{uniformity}, which designates the possible ground states of distortion,  when fulfilled, and ignites geometric frustration, when unfulfilled.

This theoretical tool has already proven useful in three-dimensional space. There, it amounts to the request that the field $\n$ appears everywhere distorted in the same way. More descriptively, this amounts to say that an observer placed everywhere in space would represent the tensorial descriptor of distortion $\nabla\n$ by use of the same (constant) scalars in a \emph{distortion} frame $\framed$ intrinsically associated with the nematic field. 

In \cite{virga:uniform}, following \cite{selinger:interpretation} and \cite{machon:umbilic}, $\nabla\n$ is represented in the form
\begin{equation}
	\label{eq:nabla_n_representation}
	\nabla\n=-\bend\otimes\n+\frac12T\W(\n)+\frac12S\Proj(\n)+\D,
\end{equation}
where $S:=\diver\n$ is the \emph{splay}, $T:=\n\cdot\curl\n$ is the \emph{twist}, $\bend:=\n\times\curl\n$ is the \emph{bend}, $\W(\n)$ is the skew-symmetric tensor with axial vector $\n$, $\Proj(\n):=\I-\n\otimes\n$ is the projector in the plane orthogonal to $\n$, and $\D$ is a symmetric tensor that annihilates $\n$, $\D\n=\zero$, and which admits the following biaxial representation,
\begin{equation}
	\label{eq:D_representation}
	\D=q(\n_1\otimes\n_1-\n_2\otimes\n_2).
\end{equation}
In \eqref{eq:D_representation}, $q>0$ and $(\n_1,\n_2)$ is a pair of orthogonal unit vectors in the plane orthogonal to $\n$, oriented so that $\n=\n_1\times\n_2$.\footnote{The orientation of both $\n_1$ and $\n_2$ can be chosen freely, without affecting $\D$, which is uniquely determined by $\nabla\n$.} The distortion frame is thus built solely from $\D$.

Letting $\bend=b_1\n_1+b_2\n_2$, we call $(S,T,b_1,b_2,q)$ the distortion \emph{characteristics} of the field $\n$ and we say that the latter is \emph{uniform} whenever the former are constant in space.\footnote{It is argued in \cite{selinger:director} that $q$ should be called the \emph{tetrahedral} splay; we would rather prefer to call it the \emph{octupolar} splay, as done in \cite{paparini:stability}, to acknowledge the role played by a cubic (octupolar) potential on the unit sphere in representing all distortion characteristics, but $T$ \cite{pedrini:liquid,gaeta:review}.}

It was proved in \cite{virga:uniform} that the most general uniform nematic field of class $C^2$ in a flat three-dimensional space is characterized by the conditions,
\begin{equation}
	\label{eq:uniuformity_3_D}
	S=0,\quad T =\pm2q,\quad b_1 =\pm b_2 =b,
\end{equation}
where $q>0$ and $b$ are arbitrary scalar parameter. This precisely transliterates
Meyer's \emph{heliconical} distortion \cite{meyer:structural}, which was observed experimentally in the ground state of \emph{twist-bend} nematic phases \cite{cestari:phase,borshch:nematic}. A similar result, perhaps geometrically more attractive, has been proved for \emph{curved} three-dimensional spaces in \cite{pollard:intrinsic,dasilva:moving}.

On a surface $\surface$ the picture changes considerably and the very definition of uniform nematic field $\n$ must be rethought about. Usually, the covariant gradient $\nablac\n$ is employed as a local tensorial descriptor of distortion. We shall follow an alternative avenue, which in Sec.~\ref{sec:uniform}, after pausing on a number of mathematical preliminaries in Sec.~\ref{sec:prelim}, will lead us to an equivalent definition of surface uniformity.

A cornerstone in this field is Niv \& Efrati's result \cite{niv:geometric} to the effect that uniform nematic fields are only possible on surfaces with constant \emph{negative} Gaussian curvature (whose value is dictated by the appropriate distortion characteristics). Apart from this necessary condition, nothing else is known in the literature about uniform fields on surfaces. In Sec.~\ref{sec:parallel}, we show that the existence of these fields is tightly linked to the existence of special systems of \emph{uniform} geodesics of $\surface$ that  parallel transport $\n$ (in Levi-Civita's sense). We prove that, for every geodesic $\Curve$ of $\surface$, there are \emph{two} such systems that include $\Curve$. Such a duality propagates to the conveyed uniform fields, in a way reminiscent of the alternative summarized in \eqref{eq:uniuformity_3_D}.

In Sec.~\ref{sec:pseudosphere}, we construct explicitly all uniform geodesics and conveyed uniform fields for Beltrami's pseudosphere. Since, by Minding's theorem, all surfaces with constant Gaussian curvature are isometric, the  solution found for the pseudosphere can (in principle) be extended to all surfaces where uniform fields exist.

Finally, in Sec.~\ref{sec:conclusions}, we collect our conclusions and some thoughts for further studies. The paper is completed by two Appendices with auxiliary results and an animation showing a typical uniform field on the pseudosphere.

\section{Mathematical preliminaries}\label{sec:prelim}
In this section, to make our development self-contained and to set forth the notation employed throughout the paper,  we recall a few preliminary results about calculus on surfaces.\footnote{The experienced reader would likely jump ahead to the following section.} We shall use an \emph{absolute} approach to surface calculus,\footnote{That is, not relying on coordinates, as we abhor indices at least as much as Cartan did, see \cite[p.\,xvii]{needham:visual}} generally inspired by the work of Weatherburn \cite{weatherburn:differential_1,weatherburn:differential_2}, whose essential features are also succinctly outlined in \cite{sonnet:bending-neutral}.\footnote{The work of Weatherburn was preceded by the introduction of a general vector method in Differential Geometry by the Italian school that originated from Levi-Civita (see \cite{burali-forti:fondamenti,burgatti:teoremi,burgatti:memorie,burgatti:analisi} for the relevant historical sources and \cite{rosso:parallel} for a more recent application to soft matter science.)}

Let $\surface$ be a smooth (at least of class $C^2$), orientable surface imbedded in three-dimensional space $\euclid$. We denote by $\normal$ one orientation of the unit normal on $\surface$.

A major role is played below by the notion of \emph{surface gradient}, which we introduce with the aid of smooth curves $\curve(t)$ on $\surface$. A scalar field $\varphi:\surface\to\mathbb{R}$ is differentiable on $\surface$ whenever we can write
\begin{equation}
	\label{eq:differentiability_scalar}
	\frac{\dd}{\dd t}\varphi(\curve(t))=\nablas\varphi\cdot\dot{\curve},
\end{equation}
where a superimposed dot $\dot{\null}$ denotes differentiation with respect to the parameter $t$ and $\dot{\curve}$ is a  vector along the tangent to $\curve(t)$. Similarly, for a vector field $\vv:\surface\to\transl$, where $\transl$ is the translation space associated with $\euclid$,\footnote{Our notation for $\euclid$ and $\transl$ is the same as in \cite[p.\,324]{truesdell:first}, where these geometric structures are further illuminated, especially in connection with their role in formulating  modern continuum mechanics.}
\begin{equation}
	\label{eq:differentiation_vector}
	\frac{\dd}{\dd t}\vv(\curve(t))=(\nablas\vv)\dot{\curve},
\end{equation}
where the second-rank tensor $\nablas\vv$, which annihilates the normal $\normal$, is the surface gradient of $\vv$. In particular, $\nablas\normal$ is the \emph{curvature tensor} of $\surface$: it is a symmetric tensor, whose eigenvalues are the \emph{principal} curvatures of $\surface$.

Letting $\proj:=\I-\normal\otimes\normal$ represent the projection onto the local tangent plane $\tplane$,\footnote{Here, the \emph{dyadic} product $\av_1\otimes\av_2$ of vectors $\av_1$ and $\av_2$ is defined as the second-rank tensor whose action on a generic vector $\uv$ is specified as $(\av_1\otimes\av_2)\uv=(\av_2\cdot\uv)\av_1$, where $\cdot$ denotes the inner product in $\transl$.} we define the \emph{covariant} gradient $\nablac\vv$ as
\begin{equation}
	\label{eq:covariant_gradient_definition}
	\nablac\vv:=\proj\nablas\vv.
\end{equation}
It reduces $\nablas\vv$ to an action onto $\tplane$, and so it can be considered as being totally inherent to the surface (or intrinsic), insensitive to the surrounding space (see also \cite[p.\,240]{needham:visual}).

The surface curl of $\vv$, $\curls\vv$, is defined by the identity
\begin{equation}
	\label{eq:surface_curl}
	2\skw(\nablas\vv)\uv=[\nablas\vv-(\nablas\vv)\trans]\uv=\curls\vv\times\uv\quad\forall\ \uv\in\transl,
\end{equation}
where $\skw$ denotes the skew-symmetric part of a second-rank tensor, a superscript $\trans$ the transposition of a tensor, and $\times$ the vector product in $\transl$. Equation \eqref{eq:surface_curl} simply says that $\curls\vv$ is the axial vector associated with $2\skw(\nablas\vv)$. In a parallel way, we define the \emph{covariant} curl of $\vv$, $\curlc\vv$, as the axial vector associated with $2\skw(\nablac\vv)$. Likewise, the surface and \emph{covariant} divergence of $\vv$ are defined as
\begin{equation}
	\label{eq:surface_covariant_divergence}
	\divs\vv:=\tr(\nablas\vv)\quad\text{and}\quad\divc\vv:=\tr(\nablac\vv),
\end{equation}
respectively, where $\tr$ denotes the trace of a second-rank tensor.

As also recalled in \cite{sonnet:bending-neutral}, if $\h$ is a \emph{tangential} vector field, that is, such that $\h\cdot\normal\equiv0$, there exists a scalar field $\varphi$ on $\surface$ such that $\h=\nablas\varphi$, if and only if
\begin{equation}
	\label{eq:integrability_scalar}
	\skw(\nablas\h)=\skw(\curvature\h\otimes\normal).
\end{equation}
Similarly, letting a second-rank tensor field $\H$ be defined on $\surface$ so that $\H\normal\equiv\zero$, there exists a vector field $\vv$ on $\surface$ such that $\H=\nablas\vv$, if and only if
\begin{equation}
	\label{eq:integrability_vector}
	\skw(\nablas\H)=\skw(\H\curvature\otimes\normal),
\end{equation}
where $\nablas\H$ is a third-rank tensor and  skw acts as follows on its generic \emph{triadic} component $\av_1\otimes\av_2\otimes\av_3$,
\begin{equation}
	\label{eq:skw_third_rank_tensor}
	2\skw(\av_1\otimes\av_2\otimes\av_3):=\av_1\otimes\av_2\otimes\av_3-\av_1\otimes\av_3\otimes\av_2.
\end{equation}

We shall employ the method of \emph{moving frames}.\footnote{This method was first introduced by Cartan \cite{cartan:methode} and is extensively used in the book \cite{o'neill:elementary} to illustrate the differential geometry of spaces, surfaces, and curves. A more recent, rather comprehensive account, far more formal than needed here, can be found in the book \cite{clelland:from}.} An orthonormal frame $\frameen$, where $\normal=\e_1\times\e_2$,\footnote{We shall only consider \emph{positively} oriented, orthonormal frames with one unit vector coincident with $\normal$.} glides over $\surface$ according to the laws
\begin{equation}\label{eq:gliding_laws}
	\begin{cases}
		\nablas\e_1=\e_2\otimes\cv+\normal\otimes\dv_1,\\
		\nablas\e_2=-\e_1\otimes\cv+\normal\otimes\dv_2,\\
		\nablas\normal=-\e_1\otimes\dv_1-\e_2\otimes\dv_2,
	\end{cases}
\end{equation}
where the vector fields $(\cv,\dv_1,\dv_2)$ are everywhere tangent to $\surface$; these are the \emph{connectors} of the moving frame. More precisely, $\cv$ is the \emph{spin} connector and $\dv_1$, $\dv_2$ are the \emph{curvature} connectors.\footnote{We call them connectors because they connect the frame at one point to the frame in a nearby point. They are further expounded in \cite{ozenda:blend,ozenda:kirchhoff,sonnet:bending-neutral}, where they are erroneously referred to as \emph{Cartesian} (instead of \emph{Cartanian}). See also \cite{sonnet:variational} for a recent application to shell theory.} Since the curvature tensor $\nablas\normal$ is symmetric, the curvature connectors must obey the identity
\begin{equation}
	\label{eq:connector_identity}
	\dv_1\cdot\e_2=\dv_2\cdot\e_1.
\end{equation}
In particular, the third equation in \eqref{eq:gliding_laws} implies that
\begin{subequations}\label{eq:mean_and_gaussian}
	\begin{align}
	2H:=&\tr\curvature=-(\dv_1\cdot\e_1+\dv_2\cdot\e_2)\label{eq:mean_curvature},\\
	K:=&\det\curvature=\dv_1\times\dv_2\cdot\normal,\label{eq:gaussian_curvature}
\end{align}
\end{subequations}
where $H$ and $K$ are the \emph{mean} and \emph{Gaussian} curvatures of $\surface$, respectively.\footnote{By $\tr\curvature$ and $\det\curvature$, we mean the sum and the product of the principal curvatures of $\surface$, respectively.}
Below, we shall apply \eqref{eq:gliding_laws} to a variety of different moving frames, each entailing its own set of connectors.

Given a smooth curve $\Curve$ on $\surface$, let $\tangent$ be the unit tangent vector to $\Curve$. The moving frame $\framet$, where $\tperp:=\normal\times\tangent$, is the \emph{Darboux} frame of $\Curve$; it satisfies the equations
\begin{equation}
	\label{eq:Darboux_frame}
	\begin{cases}
		\tangent'=\kg\tperp+\kn\normal,\\
		\tperp'=-\kg\tangent-\tg\normal,\\
		\normal'=-\kn\tangent+\tg\tperp,
	\end{cases}
\end{equation}
where a prime $'$ denotes differentiation with respect to the arc-length coordinate along $\Curve$, $\kg$ is the \emph{geodesic} curvature, $\kn$ the \emph{normal} curvature, and $\tg$ the \emph{geodesic} torsion of $\Curve$.\footnote{Equations \eqref{eq:Darboux_frame} differ by the the sign of $\tg$ from the traditional form presented in most textbooks on Differential Geometry (see, for example, \cite[p.\,264]{doCarmo:differential}).} Equations \eqref{eq:Darboux_frame} can be given a more concise (and perhaps more telling) form,
\begin{equation}
	\label{eq:Darboux_spin}
	\tangent'=\spin\times\tangent,\quad\tperp'=\spin\times\tperp,\quad\normal'=\spin\times\normal,
\end{equation}
by introducing the \emph{spin vector} $\spin$ defined as 
\begin{equation}
	\label{eq:spin_vector}
	\spin:=-\tg\tangent-\kn\tperp+\kg\normal.
\end{equation}

By applying \eqref{eq:gliding_laws} to the frame $\framet$, assuming that $\Curve$ is a field line with tangent $\tangent$ that crosses everywhere at right angles a field line $\Curve_\perp$ with tangent $\tperp$,  we can write
\begin{subequations}
	\label{eq:t_prime_t_perp_prime}
	\begin{align}
		\tangent'&=(\nablas\tangent)\tangent=(\cv\cdot\tangent)\tperp+(\dv_1\cdot\tangent)\normal,\label{eq:t_prime}\\
		\tperp'&=(\nablas\tperp)\tperp=-(\cv\cdot\tperp)\tangent+(\dv_2\cdot\tperp)\normal.\label{eq:t_perp_prime}
	\end{align}
\end{subequations}
Combining these equations with \eqref{eq:Darboux_frame}, and treating in exactly the same way the Darboux frame $(\tperp,-\tangent,\normal)$ associated with $\Curve_\perp$, characterized by the differential measures $\kgp$, $\knp$, and $\tgp$, we arrive at the following representation for the connectors,\footnote{Since by \eqref{eq:spin_vector} $\kg$ is the component of $\spin$ along $\normal$, \eqref{eq:connector_representation_c} fully justifies calling $\cv$ the \emph{spin} connector.}
\begin{subequations}
	\label{eq:connector_representation}
	\begin{align}
		\cv&=\kg\tangent+\kgp\tperp,\label{eq:connector_representation_c}\\
		\dv_1&=\kn\tangent+\tgp\tperp,\label{eq:connector_representation_d_1}\\
		\dv_2&=-\tg\tangent+\knp\tperp.\label{eq:connector_representation_d_2}
	\end{align}
\end{subequations}
For \eqref{eq:connector_identity} to be valid (with $\e_1=\tangent$ and $\e_2=\tperp$), it must then be 
\begin{equation}\label{eq:geodesic_torsion_identity}
	\tg+\tgp=0,
\end{equation}
which is a known identity (see, for example, \cite[p.\,488]{doCarmo:differential}).

For a unit vector $\n(s)$ tangent to $\surface$ prescribed along $\Curve$ as a function of $s$, we say that it is \emph{parallel transported} along $\Curve$ if
\begin{equation}
	\label{eq:parallel_transport_definition}
	\n'=\spin_\parallel\times\n,
\end{equation}
where the spin vector $\spin_\parallel$ is such that $\spin_\parallel\cdot\normal\equiv0$.\footnote{This definition of parallel transport along a curve is that of Levi-Civita~\cite{levi-civita:nozione}, as interpreted in the kinematic analogy of Persico~\cite{persico:realizzazione}. $\spin_\parallel$ can also be characterized as the \emph{least} spin vector $\spin$ that makes $\n$ glide along $\Curve$ as part of the moving frame $\framen$, see \cite{rosso:parallel}.} Equivalently, \eqref{eq:parallel_transport_definition} prescribes $\n'$ to be parallel to $\normal$. Letting $\tangent$ be the unit vector tangent to $\Curve$, we can represent $\n$ as
\begin{equation}
	\label{eq:n_alpha_representation}
	\n=\cos\gamma\tangent+\sin\gamma\tperp,
\end{equation}
where $\gamma=\gamma(s)$ is a smooth function. With the aid of \eqref{eq:Darboux_frame}, we can easily compute
\begin{equation}
	\label{eq:n_prime}
	\n'=(\gamma'+\kg)\nperp+(\kn\cos\gamma-\tg\sin\gamma)\normal,
\end{equation}
which shows that $\n$ is parallel transported along $\Curve$ whenever
\begin{equation}
	\label{eq:parallel_transport_equation}
	\gamma'+\kg=0.
\end{equation}
In particular, if $\kg\equiv0$, which is the case if $\Curve$ is a geodesic of $\surface$, then $\gamma$ is constant along $\Curve$.

A surface $\surface$ can also be represented by use of coordinates $(u,v)$ as the image of a mapping $\rv:\Omega\to\euclid$, where $\Omega$ is a domain in $\mathbb{R}^2$. For a surface of class $C^2$, coordinates $(u,v)$ can be chosen so as to be, at least locally, \emph{isothermal},\footnote{For surfaces of class $C^1$, isothermal coordinates can fail to exist. The existence of isothermanl coordinates for $C^2$ surfaces is established by a classical theorem revisited in  \cite{chern:elementary}.} that is, such that 
\begin{equation}
	\label{eq:isothermal_coordinates}
	\rvu\cdot\rvv=0\quad\text{and}\quad|\rvu|=|\rvv|,
\end{equation}
where $\rvu:=\partial_u\rv$ and $\rvv:=\partial_v\rv$. Letting 
\begin{equation}
	\label{eq:e_u_e_v_definition}
	\e_u:=\frac{\rvu}{|\rvu|}\quad\text{and}\quad\e_v:=\frac{\rvv}{|\rvv|},
\end{equation}
we orient $\surface$ so that $\normal=\e_u\times\e_v$. The moving frame $\framee$ is subject to \eqref{eq:gliding_laws}, where the appropriate connectors will be denoted as $(\cv,\dv_u,\dv_v)$.

In the following section, we shall start our study by defining when a unit vector field $\n$ tangent to $\surface$ is said to be \emph{uniform}.

\section{Surface uniformity}\label{sec:uniform}
Nematic uniformity has been tackled already in three-dimensional spaces, either flat or curved, where this notion is fairly well understood. For surfaces imbedded in $\euclid$, we need first agree on a definition of uniformity for a tangential unit field $\n$. To this end, we assume that a tangent unit vector field $\n$ of class at least $C^2$ is prescribed on $\surface$ and we write the gliding laws in \eqref{eq:gliding_laws} for the moving frame $\framen$, where $\nperp$ is chosen so that $\normal=\n\times\nperp$,
\begin{equation}\label{eq:gliding_laws_n}
	\begin{cases}
		\nablas\n=\nperp\otimes\cv+\normal\otimes\dv_1,\\
		\nablas\nperp=-\n\otimes\cv+\normal\otimes\dv_2,\\
		\nablas\normal=-\n\otimes\dv_1-\nperp\otimes\dv_2.
	\end{cases}
\end{equation}
We further specialize \eqref{eq:connector_identity} in the form
\begin{equation}
	\label{eq:connector_identity_n}
	\dv_1\cdot\nperp=\dv_2\cdot\n,
\end{equation}
which guarantees the symmetry of $\nablas\normal$. This leads us to realize that, since $\dv_1=-\curvature\n$, the \emph{intrinsic} surface distortion of $\n$, as perceived by a two-dimensional observer insensitive to the way the normal to the surface $\surface$ changes in space,\footnote{Often, are called \emph{intrinsic} the properties of a surface that are independent of its imbedding in three-dimensional space, and \emph{extrinsic} those properties that depend on that. As effectively put in \cite[p.\,11]{needham:visual},
	\begin{quote}
		[Intrinsic geometry] means the geometry that is knowable to tiny, ant-like, intelligent (but 2-dimensional!) creatures living \emph{within} the surface. 
	\end{quote}
Clearly, whenever the normal to the surface features in one of its properties, that is likely to be extrinsic.} is only captured by $\cv$, or equivalently by
\begin{equation}
	\label{eq:covariant_gradient_n}
	\nablac\n=\nperp\otimes\cv.
\end{equation}
Alternatively, we may say that the knowledge of $\n$ and the full curvature tensor $\nablas\normal$ at a point of $\surface$ determines the non-covariant component of $\nablas\n$ at that point, thus revealing its \emph{extrinsic} nature.\footnote{The reader should not be led to think that the curvature tensor cannot covey intrinsic features. The well-knwon \emph{theorema egregium} of Gauss indeed shows that the Gaussian curvature $K$, being an \emph{isometric} invariant (and so depending only on the metric of the imbedded surface), is an intrinsic measure associated with $\nablas\normal$ (see, for example, \cite[p.\,291]{o'neill:elementary}). By \eqref{eq:gaussian_curvature}, this also implies that $\dv_1\times\dv_2\cdot\normal$ is an intrinsic scalar. Said differently, the extrinsic curvature $K$ of $\surface$ belongs to the intrinsic geometry of $\surface$, and can accordingly also be measured as the \emph{angular excess} per unit area \cite[p.\,140--142]{needham:visual}.}

Letting $\cv=c_1\n+c_2\nperp$, it easily follows from the first equation in \eqref{eq:gliding_laws_n} and \eqref{eq:covariant_gradient_n} that
\begin{subequations}\label{eq:c_1_c_2}
\begin{align}
	c_1&=-b_\perp:=-\n\times\curls\n\cdot\nperp=\curls\n\cdot\normal=\curlc\n\cdot\normal,\label{eq:c_1}\\
	c_2&=S:=\divs\n=\divc\n,\label{eq:c_2}
\end{align}
\end{subequations}
which identify $c_1$ with the intrinsic component of the \emph{bend} vector $\bend=\n\times\curls\n$ (to within a sign) and $c_2$ with the \emph{splay} $S$, which is fully intrinsic. A surface nematic field $\n$ has no \emph{twist}, as $\n\cdot\curls\n=\dv_1\cdot\nperp=\dv_2\cdot\n$ is purely extrinsic and, by \eqref{eq:covariant_gradient_n}, $T=\n\cdot\curlc\n\equiv0$.\footnote{Here we are guilty of some abuse of notation, as we still denote  the bend, splay, and twist of a surface field by the same symbols used for three-dimensional fields. No confusion should however arise, as here we are only concerned with surface fields.}

Thus, we shall say that a nematic surface field $\n$ tangent to $\surface$ is \emph{uniform}, whenever the spin connector $\cv$ has constant components along both $\n$ and $\nperp$, which by \eqref{eq:c_1_c_2} can be interpreted as bend and splay covarariant distortions of the field, respectively. Although formulated in the language of connectors, this definition is equivalent to the standard one formulated in terms of the covariant gradient.

We want to show that the notion of surface uniformity thus introduced is invariant under isometric deformations of the surface $\surface$, as is expected from any intrinsic property.

To this end, we recall that a deformation $\y:\surface\to\euclid$, which is  of class $C^2$ and produces the surface $\surface^\ast=\y(\surface)$ as image of $\surface$, is an isometry whenever its surface gradient can be represented as
\begin{equation}
	\label{eq:isometry_gradient}
	\nablas\y=\R\proj,
\end{equation}
where $\R$ is a member of the special orthogonal group $\orth$.\footnote{Equation \eqref{eq:isometry_gradient} is a special form of the polar decomposition theorem for the deformation of surfaces proved in \cite{man:coordinate}: it requires that the surface stretching tensor $\U$ be the projection $\proj$.} Suppose that $\n$ is a uniform field on $\surface$ and let $\n^\ast$ be the unit vector field conveyed on $\surface^\ast$ by $\y$, so that $\n^\ast=(\nablas\y)\n$. Thus, $\R$ can be represented as
\begin{equation}
	\label{eq:R_representation}
	\R=\n^\ast\otimes\n+\nperp^\ast\otimes\nperp+\normal^\ast\otimes\normal,
\end{equation}
where $\nperp^\ast=\normal^\ast\times\n^\ast$ and $\normal^\ast$ is the unit normal to $\surface^\ast$ so that $\framenast$ is oriented as $\framen$. The moving frame $\framenast$ obeys on $\surface^\ast$ the laws
\begin{equation}\label{eq:gliding_laws_n_ast}
	\begin{cases}
		\nablast\n^\ast=\nperp^\ast\otimes\cv^\ast+\normal^\ast\otimes\dv_1^\ast,\\
		\nablast\nperp^\ast=-\n^\ast\otimes\cv^\ast+\normal^\ast\otimes\dv_2^\ast,\\
		\nablast\normal^\ast=-\n^\ast\otimes\dv_1^\ast-\nperp^\ast\otimes\dv_2^\ast,
	\end{cases}
\end{equation}
where $\nablast$ denotes the surface gradient on $\surface^\ast$ and the connectors $(\dv_1^\ast,\dv_2^\ast)$ satisfy
\begin{equation}
		\label{eq:connector_identity_n_ast}
	\dv_1^\ast\cdot\nperp^\ast=\dv_2^\ast\cdot\n^\ast,
\end{equation}
which like \eqref{eq:connector_identity_n} ensures that $\nablast\normal^\ast$ is symmetric.

Our aim is to prove that $\cv^\ast$ has constant components in the frame $\framenast$
whenever $\cv$ has constant components in the frame $\framen$. We shall reach this conclusion  as a consequence of the requirement of integrability in \eqref{eq:integrability_vector} applied to \eqref{eq:isometry_gradient}.

Setting 
\begin{equation}
	\label{eq:H_definition}
	\H:=\nablas\y=\n^\ast\otimes\n+\nperp^\ast\otimes\nperp,
\end{equation}
we first compute
\begin{equation}
	\label{eq:H_integrability_left}
	\begin{split}
		\nablas\H&=\n^\ast\otimes[\nperp\otimes\proj(\cv-\R\trans\cv^\ast)+\normal\otimes\dv_1]+\nperp^\ast\otimes[\n\otimes\proj(\R\trans\cv^\ast-\cv)+\normal\otimes\dv_2]\\
		&+\normal^\ast\otimes[\n\otimes\proj\R\trans\dv_1^\ast+\nperp\otimes\proj\R\trans\dv_2^\ast],
	\end{split}
\end{equation}
where we have used the identity
\begin{eqnarray}
	\label{eq:chain_rule}
	\nablas\uv^\ast=\nablast\uv^\ast\R\proj,
\end{eqnarray}
valid, by the chain rule, for a generic differentiable vector field $\uv^\ast$ on $\surface^\ast$, and we have applied several times both \eqref{eq:gliding_laws_n} and \eqref{eq:gliding_laws_n_ast}. Similarly, we compute 
\begin{equation}
	\label{eq:H_integrability_right}
	\H\curvature\otimes\normal=-\n^\ast\otimes\dv_1\otimes\normal-\nperp^\ast\otimes\dv_2\otimes\normal.
\end{equation}
Now, requiring \eqref{eq:integrability_vector} to hold amounts to require that three skew-symmetric tensors vanish, one for each left component of $\nablas\H$ in the frame $\framenast$. These identities eventually require that the corresponding axial vectors vanish. The integrability condition is thus reduced to the equations
\begin{equation}
	\label{eq:integrability_condition_reduced}
	\n\times\proj(\cv-\R\trans\cv^\ast)=\zero,\quad\nperp\times\proj(\cv-\R\trans\cv^\ast)=\zero,\quad \nperp\cdot\R\trans\dv_1^\ast=\n\cdot\R\trans\dv_2^\ast.
\end{equation}
While the last one, by \eqref{eq:R_representation}, is equivalent to \eqref{eq:connector_identity_n_ast}, and so it is automatically satisfied, the first two are equivalent to
\begin{equation}
	\label{eq:c_c_ast_relation}
	\cv-\R\trans\cv^\ast=\lambda\normal,
\end{equation}
for some scalar $\lambda$. Since $\cv\cdot\normal=0$, \eqref{eq:c_c_ast_relation} and \eqref{eq:R_representation} imply that $\lambda=\cv^\ast\cdot\normal^\ast=0$, and so
\begin{equation}
	\label{eq:c_ast=R_c}
	\cv^\ast=\R\cv=-b_\perp\n^\ast+S\nperp^\ast,
\end{equation}
which proves that the field $\n^\ast$ is uniform with  the same components of bend and splay as $\n$. Thus, as desired, we conclude that the notion of uniformity is isometrcally invariant: a field uniform on a surface $\surface$ will be conveyed into a field uniform on all surfaces isometric to $\surface$.

In Appendix~\ref{sec:conditions}, we  characterize the connectors of a uniform field; in particular, it is shown that only surfaces with a constant \emph{negative} Gaussian curvature $K$ can bear uniform fields, a result already known from \cite{niv:geometric}. In the following section, we shall see how another necessary condition for surface uniformity leads us to identify all possible uniform fields.

\section{Uniform Geodesics}\label{sec:parallel}
As shown in \cite{niv:geometric}, a surface $\surface$ can host a uniform nematic field $\n$ only if its Gaussian curvature $K$ obeys
\begin{equation}
	\label{eq:K_necessary_condition}
	K\equiv-(b_\perp^2+S^2),
\end{equation}
where $b_\perp$ and $S$ are  the bend and splay (constant) \emph{distortion components} of the field. Surfaces with constant negative Gaussian curvature are called \emph{pseudospherical}, following the use of Beltrami, who found their simplest exemplar, the \emph{pseudosphere} (see, for example, \cite[p.\,22]{needham:visual}). 
Henceforth, to simply our development (without affecting its generality), we shall rescale lengths so that $K\equiv-1$. Accordingly, by \eqref{eq:c_1_c_2}, we write $\cv$ as
\begin{equation}
	\label{eq:c_representation}
	\cv=-\sin\alpha\n+\cos\alpha\nperp,\quad\text{where}\quad\cos\alpha=S\quad\text{and}\quad\sin\alpha=b_\perp,
\end{equation}
so that 
\begin{equation}
	\label{eq:distortion_anisotropy_definition}
	\tan\alpha=\frac{b_\perp}{S},
\end{equation}
which will henceforth be referred to as the \emph{distortion anisotropy}.

Let a uniform field $\n$ be prescribed on a pseudospherical surface $\surface$. We seek a curve $\Curve$ that parallel transport $\n$ starting from an arbitrary point $\x\in\surface$. If $\tangent$ denotes the unit tangent vector to $\Curve$, by \eqref{eq:parallel_transport_definition}, it must be
\begin{equation}
	\label{eq:n_prime_n_perp=0}
	\n'\cdot\n_\perp\equiv0,
\end{equation}
where a prime denotes differentiation in the arc-length parameter along $\Curve$. Since $\n'=(\nablas\n)\tangent$, by \eqref{eq:gliding_laws_n}, \eqref{eq:n_prime_n_perp=0} is just the same as
\begin{equation}
	\label{eq:t_c_=0}
	\tangent\cdot\cv\equiv0,
\end{equation}
which, by \eqref{eq:c_representation}, amounts to choose
\begin{equation}
	\label{eq:t_choices}
	\tangent=\pm(\cos\alpha\n+\sin\alpha\nperp).
\end{equation}
By differentiating both sides of \eqref{eq:t_choices} along $\Curve$, again by use of \eqref{eq:gliding_laws_n}, since $\alpha$ is constant, we arrive at
\begin{equation}
	\label{eq:t_choices_prime}
	\tangent'=\pm\{(\dv\cdot\tangent)\cos\alpha+(\dv_2\cdot\tangent)\sin\alpha\}\normal,
\end{equation}
which shows that $\Curve$ must also be a geodesic of $\surface$. Thus, a necessary condition for $\n$ to be a uniform field is to be parallel transported along geodesics.

Letting $\tperp=\normal\times\tangent$, we readily derive from \eqref{eq:c_representation} and \eqref{eq:t_choices} that
\begin{equation}
	\label{eq:t_perp_representation}
	\tperp=\pm(-\sin\alpha\n+\cos\alpha\nperp)=\pm\cv,
\end{equation}
which in turn implies that
\begin{equation}
\label{eq:n_n_perp_representations}
\n=\pm(\cos\alpha\tangent-\sin\alpha\tperp)\quad\text{and}\quad\nperp=\pm(\sin\alpha\tangent+\cos\alpha\tperp).
\end{equation}

The angle $\alpha$ remains constant along the geodesic $\Curve$; its value is uniquely determined by the distortion components prescribed for a uniform field. Of course, a uniform field  may well fail to exist on $\surface$, but if it does exist, it must be parallel transported by a system of geodesics with relative orientation determined by the distortion components. Among all possible systems of geodesics on $\surface$, we call \emph{uniform} those that parallel transport a uniform field with constant angle $\alpha$.

There is a simple criterion to identify systems of uniform geodesics. Consider first the frame $\framet$ associated with a system of uniform geodesics.\footnote{There two such frames, corresponding to the two possible choices of sign in \eqref{eq:t_choices}.} By computing $\nablas\tangent$ from \eqref{eq:t_choices} (and, correspondingly, $\nablas\tperp$ from \eqref{eq:t_perp_representation}), we easily see that the gliding laws \eqref{eq:gliding_laws} are obeyed with connectors $(\hat{\cv},\hat{\dv}_1,\hat{\dv}_2)$ given by
\begin{equation}
	\label{eq:hat_connectors}
	\hat{\cv}=\cv,\quad\hat{\dv_1}=\pm(\cos\alpha\dv_1+\sin\alpha\dv_2),\quad \hat{\dv_2}=\pm(\cos\alpha\dv_2-\sin\alpha\dv_1),
\end{equation}
where $(\cv,\dv_1,\dv_2)$ are the connectors of the frame $\framen$.\footnote{So that, in particular, $\hat{\dv_1}\times\hat{\dv_2}=\dv_1\times\dv_2=-\normal$, as required.} Since the frames $\framet$ and $\framen$ have one and the same spin connector $\cv$, which by \eqref{eq:c_representation} has unit length, it follows from \eqref{eq:t_perp_representation} that a system of uniform  geodesics must satisfy the following condition,\footnote{It is also clear from being $\tperp\cdot(\nablas\tangent)\tperp=\tperp\cdot(\nablac\tangent)\tperp$ that \eqref{eq:uniform_geodesic_condition} is an intrinsic condition.}
\begin{equation}
	\label{eq:uniform_geodesic_condition}
	\cv\cdot\tperp=\tperp\cdot(\nablas\tangent)\tperp=\pm1.
\end{equation}
Conversely, if \eqref{eq:uniform_geodesic_condition} is valid for a system of geodesics, then, as the spin connector $\cv$ of the frame $\framet$ is such that $\cv\cdot\tangent=(\nablas\tangent)\trans\tperp\cdot\tangent=0$, it can only be written in either of the following forms,
\begin{equation}
	\label{eq:c_parallel_to_tperp}
	\cv=\pm\tperp.
\end{equation}
By applying \eqref{eq:n_n_perp_representations}, with sign chosen equal to the one featuring in \eqref{eq:c_parallel_to_tperp}, it is easy to prove that, for any constant $\alpha$, the  gliding laws of $\framet$ deliver the equations
\begin{equation}
	\label{eq:duality}
	S=\divs\n=\cos\alpha\quad\text{and}\quad-b_\perp=\curls\n\cdot\normal=\sin\alpha,
\end{equation} 
for \emph{both} signs occurring in \eqref{eq:c_parallel_to_tperp} [and \eqref{eq:n_n_perp_representations}] Thus, systems of geodesics that satisfy \eqref{eq:uniform_geodesic_condition}  with one sign or the other generate uniform fields characterized by the same distortion components $S$ and $b_\perp$ (not just the same distortion anisotropy). If, starting from a given geodesic $\Curve$ of $\surface$,  equation \eqref{eq:uniform_geodesic_condition} has a solution for each choice of sign, then we generate two systems of uniform geodesics that include $\Curve$ and convey  uniform fields with equal distortions. We would then say that this establishes  a conjugation   by \emph{duality} among uniform fields. If  existing, such a duality should not be confused with the usual nematic symmetry, which identifies both $\n$ and $-\n$ with a headless \emph{director}, although this symmetry is also involved here. In going from one uniform field to its conjugate, we need first reverse the sign of $\n$ on $\Curve$ and then apply \eqref{eq:n_n_perp_representations} to the  solution of the conjugate form of \eqref{eq:uniform_geodesic_condition} (with the sign reversed). The resulting field $\n$ is \emph{not} the opposite of its dual, as both $S$ and $b_\perp$ are preserved, instead of being changed in their opposite, as they would if $\n$ were changed into $-\n$ also away from $\Curve$.\footnote{The nematic symmetry is not essential to this reasoning: alternatively, one can reverse the orientation of $\Curve$, that is, change $\tangent$ into $-\tangent$, in going from one form of \eqref{eq:uniform_geodesic_condition} to the other (with the sign changed), so as to leave the trace of $\n$ on $\Curve$ unchanged; the same conclusions as above would follow. The only things that really matter are the \emph{directions} of $\n$ and $\tangent$.}

Thus, finding all uniform nematic fields on a  surface $\surface$ with $K\equiv-1$ amounts to find all systems of geodesics that obey \eqref{eq:uniform_geodesic_condition}. By Minding's theorem (see, for example, either \cite[p.\,336]{needham:visual} or \cite[p.\,416]{pressley:elementary}), all surfaces with equal constant Gaussian curvature are (at least locally) isometric.\footnote{The classical proof of Minding's theorem requires that $\surface$ be of class at least $C^3$; for pseudospherical surfaces of class $C^2$, Minding's theorem has been proved in \cite{dorfmeister:minding}.} Since both geodesics and uniformity are preserved by isometries, it suffices to find explicitly all systems of uniform geodesics of Beltrami's pseudosphere to find (in principle) those of all pseudospherical surfaces, and thus retrace all possible uniform nematic fields on surfaces.

\section{Pseudosphere}\label{sec:pseudosphere}
Here, we first find all geodesics on the pseudosphere and then determine those that are uniform, according to our definition above. The pseudosphere with $K\equiv-1$ can be represented by the isothermal coordinates $(u,v)$  ranging in the domain $\Omega:=[0,2\pi]\times (1,+\infty)$ by the mapping $\rv$ defined as
\begin{eqnarray}
	\label{eq:pseudosphere:r}
	\rv(u,v)=\frac{1}{v}\cos u\e_1+\frac{1}{v}\sin u\e_2+\left(\ln(v+\sqrt{v^2-1}) -\frac{\sqrt{v^2-1}}{v}\right)\e_3,
\end{eqnarray}
where $\framec$ is a Cartesian frame in three-dimensional translation space $\transl$. The coordinate moving frame $\framee$ associated with \eqref{eq:pseudosphere:r} is
\begin{subequations}
	\label{eq:pseudosphere:frame_e}
	\begin{numcases}{}
	\e_u=-\sin u\e_1+\cos u\e_2,\\
	\e_v=-\frac{1}{v}(\cos u\e_1	+\sin u\e_2)+\frac{\sqrt{v^2-1}}{v}\e_3,\\
	\normal=\frac{\sqrt{v^2-1}}{v}(\cos u\e_1+\sin u\e_2)+\frac{1}{v}\e_3,
	\end{numcases}	
\end{subequations}
and the corresponding gliding laws read as,
\begin{subequations}
	\label{eq:pseudosphere:gliding_laws}
	\begin{numcases}{}
		\nablas\e_u=\e_v\otimes\e_u-\sqrt{v^2-1}\normal\otimes\e_u,\\
		\nablas\e_v=-\e_u\otimes\e_u+\frac{1}{\sqrt{v^2-1}}\normal\otimes\e_v,\\
		\nablas\normal=\sqrt{v^2-1}\e_u\otimes\e_u-\frac{1}{\sqrt{v^2-1}}\e_v\otimes\e_v.
	\end{numcases}
\end{subequations}

\begin{figure}[h] 
		\centering
		\includegraphics[width=.33\linewidth]{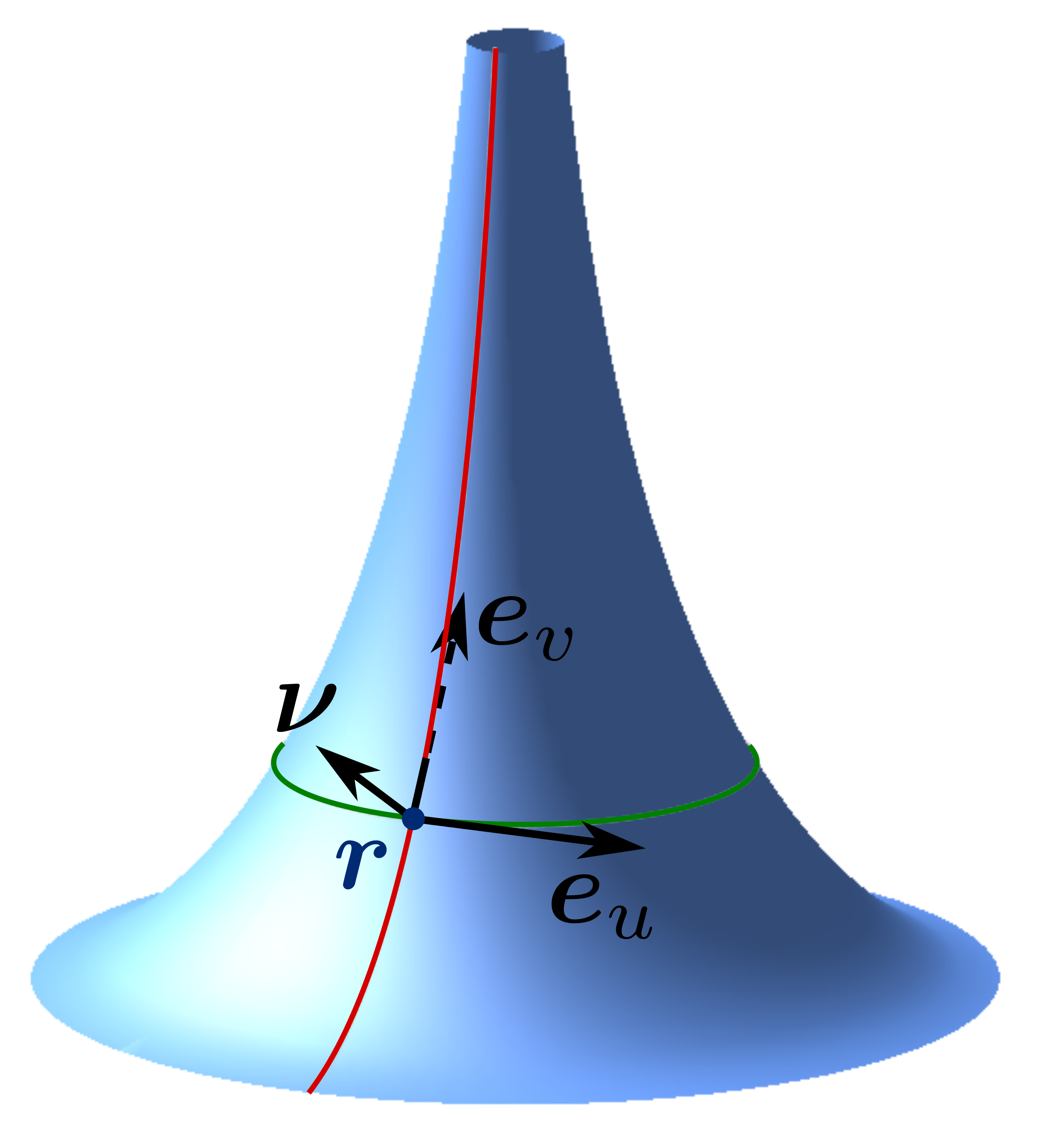}
		\caption{Beltrami's pseudosphere represented in the parametrization \eqref{eq:pseudosphere:r}. The coordinate frame $\framee$ in \eqref{eq:pseudosphere:frame_e} glides over it in accordance with \eqref{eq:pseudosphere:gliding_laws}.}
		\label{fig:pseudosphere}
\end{figure}

\begin{figure}[h] 
	\centering
	\includegraphics[width=.4\linewidth]{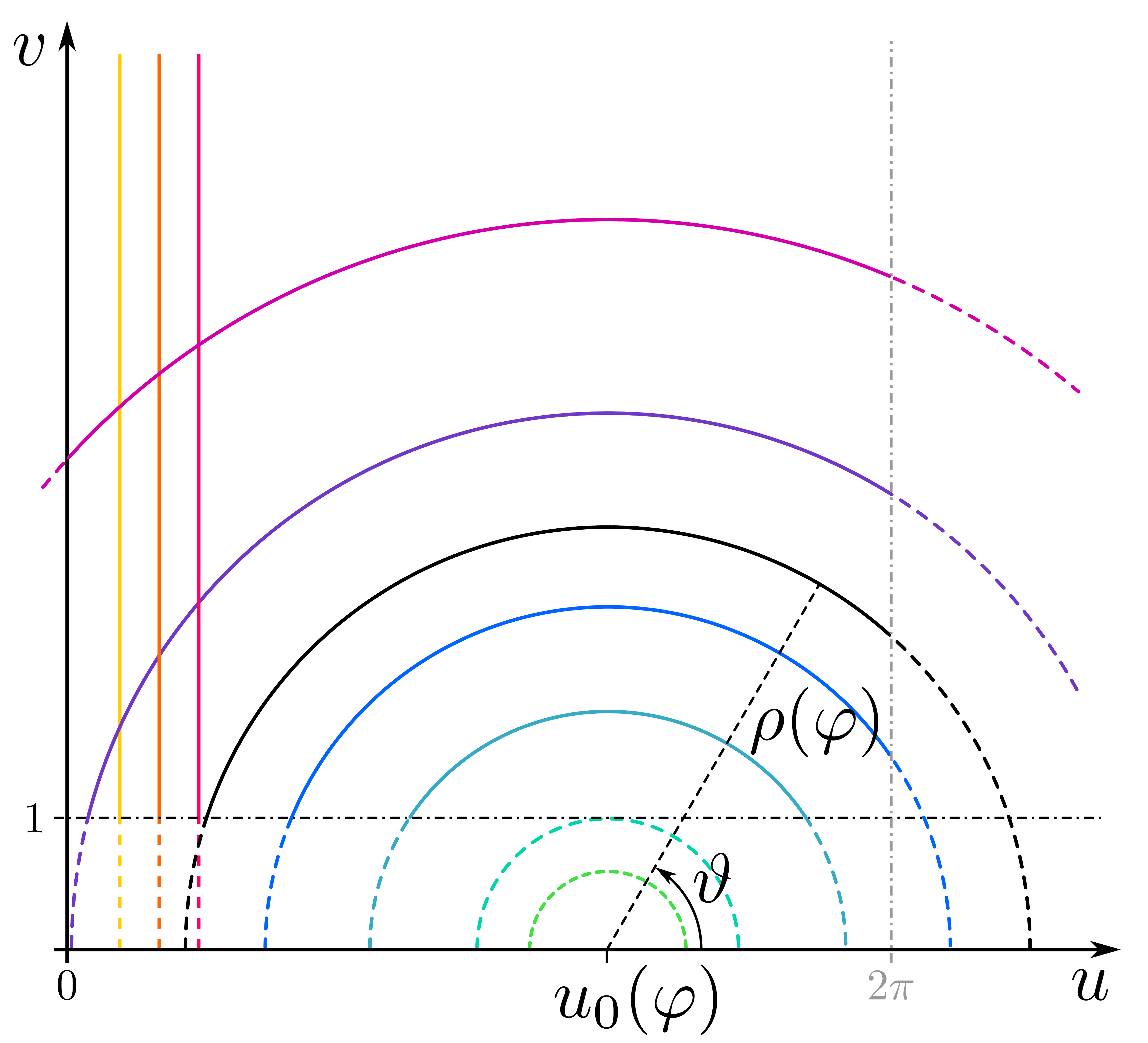}
	\caption{Pre-images of geodesics in the $(u,v)$ plane. The meridians of $\surface$ are straight lines parallel to the $v$-axis. All other geodesics are families  of circles with centers on the $u$-axis. The particular family represented here is obtained by choosing in \eqref{eq:pseudosphere_geodesic_coordinates} $u_0(\varphi)$ as a constant  and $\rho(\vp)$ as a monotonic function of $\vp$, so as to make \eqref{eq:pseudosphere_invertibility} satisfied. A single circle in this family is drawn for a fixed value of $\vp$ and $\vt$ ranging in the admissible interval \eqref{eq:pseudosphere_admissible_interval}.}
	\label{fig:pseudosphere_geodesics} 
\end{figure}

Figure~\ref{fig:pseudosphere} shows a traditional picture of the pseudosphere: it is an unbounded cylindrically-symmetric surface based on a circular rim corresponding to the segment at $v=1$ on the boundary of $\Omega$.
All geodesics can be determined by use of Clairaut's theorem (see, for example, \cite[pp.\,230-233]{pressley:elementary}): they are the family of meridians, whose pre-images in the $(u,v)$ plane are straight lines parallel to the $v$-axis, and the families of curves whose pre-images in the $(u,v)$ plane are circles with centers along the $u$-axis, see Fig.~\ref{fig:pseudosphere_geodesics}.
The whole collection of the latter are represented by
\begin{equation}
	\label{eq:pseudosphere_geodesic_coordinates}
	\begin{cases}
		u=u_0(\vp)+\rho(\vp)\cos\vt,\\
		v=\rho(\vp)\sin\vt,
	\end{cases}
\end{equation}
where $u_0(\vp)$ and $\rho(\vp)$ are smooth functions, ranging in $\mathbb{R}$ and $(1,+\infty)$, respectively, of one parameter, $\vp$, while the other parameter, $\vt$, ranges in an admissible interval (depending on $\vp$) enclosed within
\begin{equation}
\label{eq:pseudosphere_admissible_interval}
\arcsin\frac{1}{\rho}<\vt<\pi-\arcsin\frac{1}{\rho},
\end{equation}
so that $\rho(\vp)\sin\vt>1$.
A single family of geodesics is identified by taking $u_0$ constant and letting $\rho$ depend monotonically on $\vp$. Within such a family, one geodesic is singled out by fixing $\vp$ and letting $\vt$ vary in the admissible interval \eqref{eq:pseudosphere_admissible_interval}, see Fig.~\ref{fig:pseudosphere_geodesics}.

We can view \eqref{eq:pseudosphere_geodesic_coordinates} as a family of changes of variables, all expressing $(u,v)$ in terms of the \emph{geodesic} coordinates $(\vp,\vt)$; there is one such family for each choice of functions $u_0$ and $\rho$. A simple computation shows that the Jacobian of the transformation $(\vp,\vt)\mapsto(u,v)$ does not vanish whenever
\begin{equation}
	\label{eq:pseudosphere_invertibility}
	u_0'\cos\vt+\rho'\neq0,
\end{equation}
where a prime denotes differentiation with respect to $\vp$. Thus, for each choice of functions $u_0$ and $\rho$ that satisfy the (local) invertibility condition \eqref{eq:pseudosphere_invertibility}, we can cover the domain $\Omega$ with all possible  pre-images of geodesics (and $\surface$ with all possible geodesics). This will play a central role in identifying all systems of uniform geodesics of $\surface$.

First, we ask whether meridians are one such system. Since in this case we can choose $\tangent=\pm\e_v$, we easily answer the question for the \emph{positive}, as correspondingly $\tperp=\mp\e_u$ and \eqref{eq:uniform_geodesic_condition} follows as an immediate consequence of \eqref{eq:pseudosphere:gliding_laws}.

\begin{figure}[h] 
	\centering
	\begin{subfigure}[c]{0.4\linewidth}
		\centering
		\includegraphics[width=\linewidth]{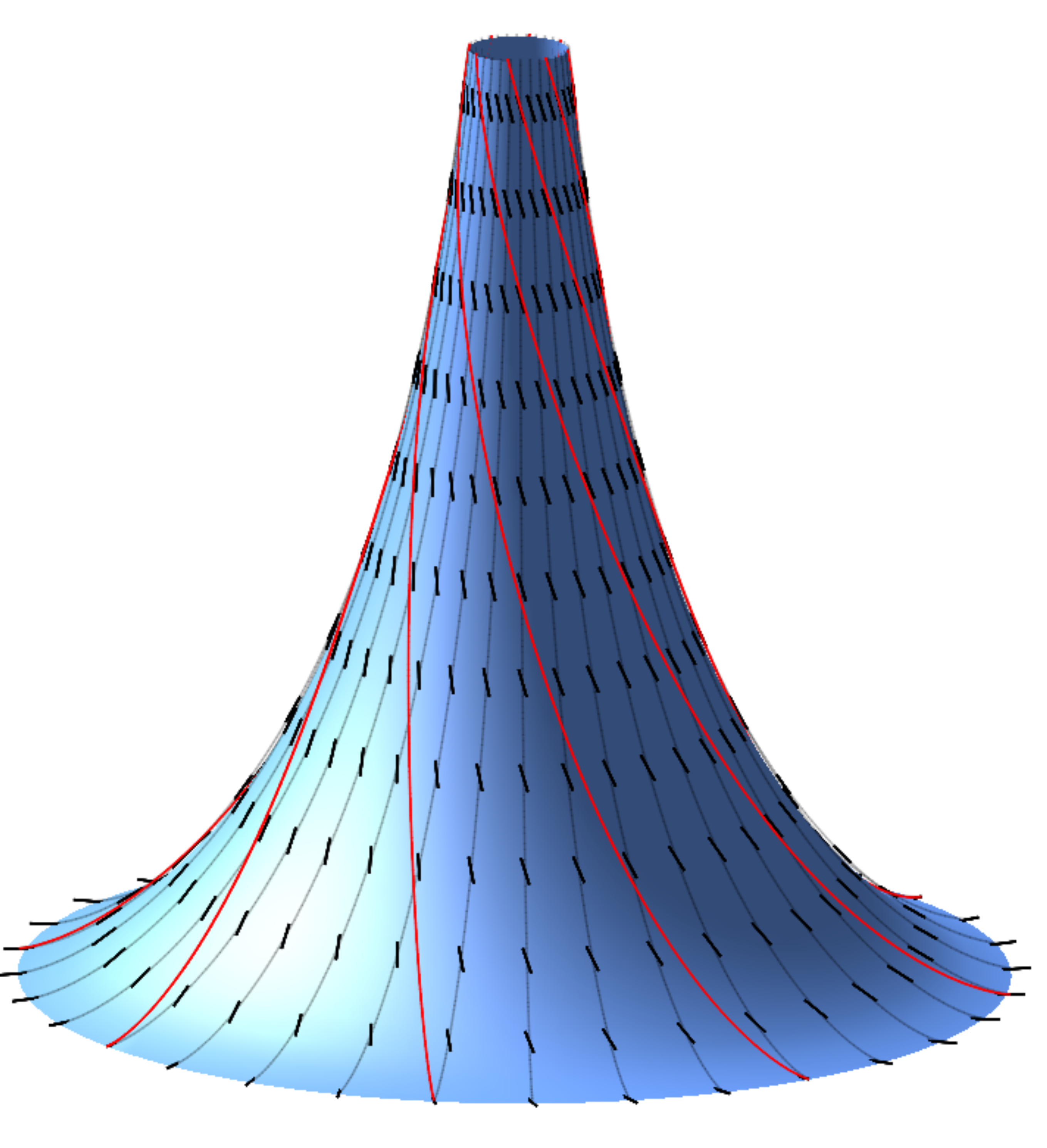}
		\caption{$\alpha=-\frac{5\pi}{12}$}
		\label{fig:meridians_a}
	\end{subfigure}
	\qquad
	\begin{subfigure}[c]{0.4\linewidth}
		\centering
		\includegraphics[width=\linewidth]{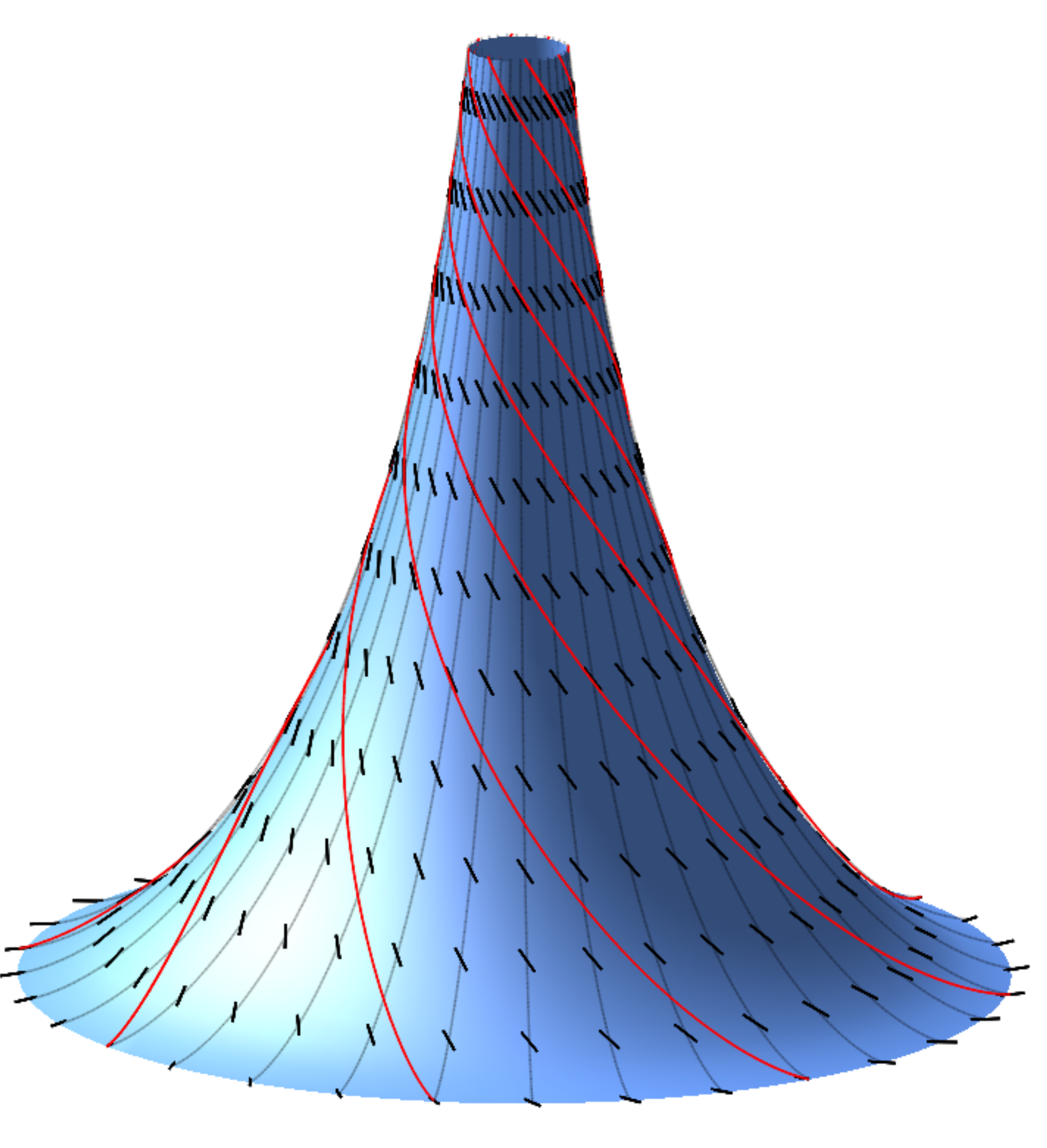}
		\caption{$\alpha=-\frac{\pi}{3}$}
		\label{fig:meridians_b}
	\end{subfigure}\\
\begin{subfigure}[c]{0.4\linewidth}
	\centering
	\includegraphics[width=\linewidth]{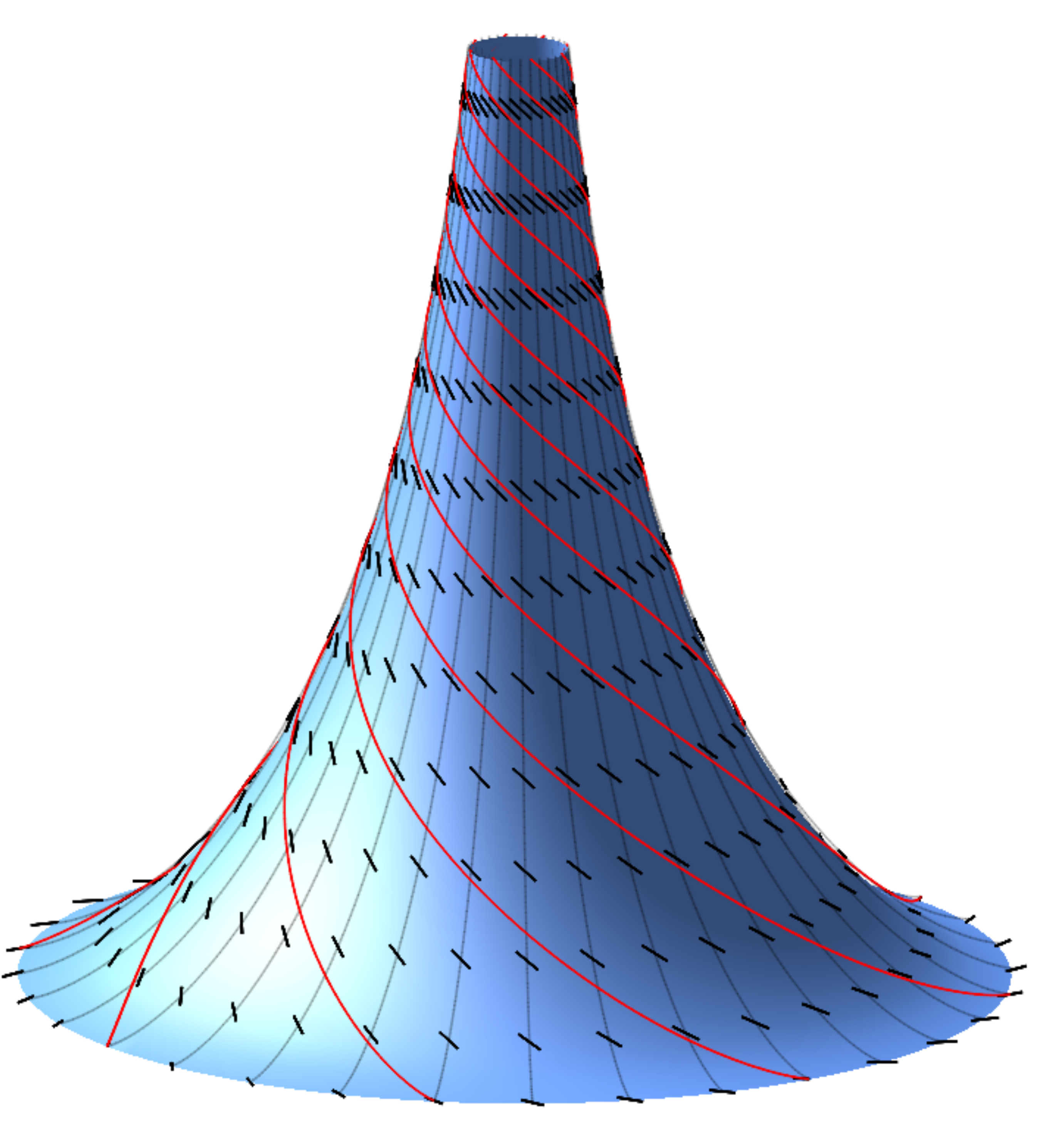}
	\caption{$\alpha=-\frac{\pi}{4}$}
	\label{fig:meridians_c}
\end{subfigure}
\qquad
\begin{subfigure}[c]{0.4\linewidth}
	\centering
	\includegraphics[width=\linewidth]{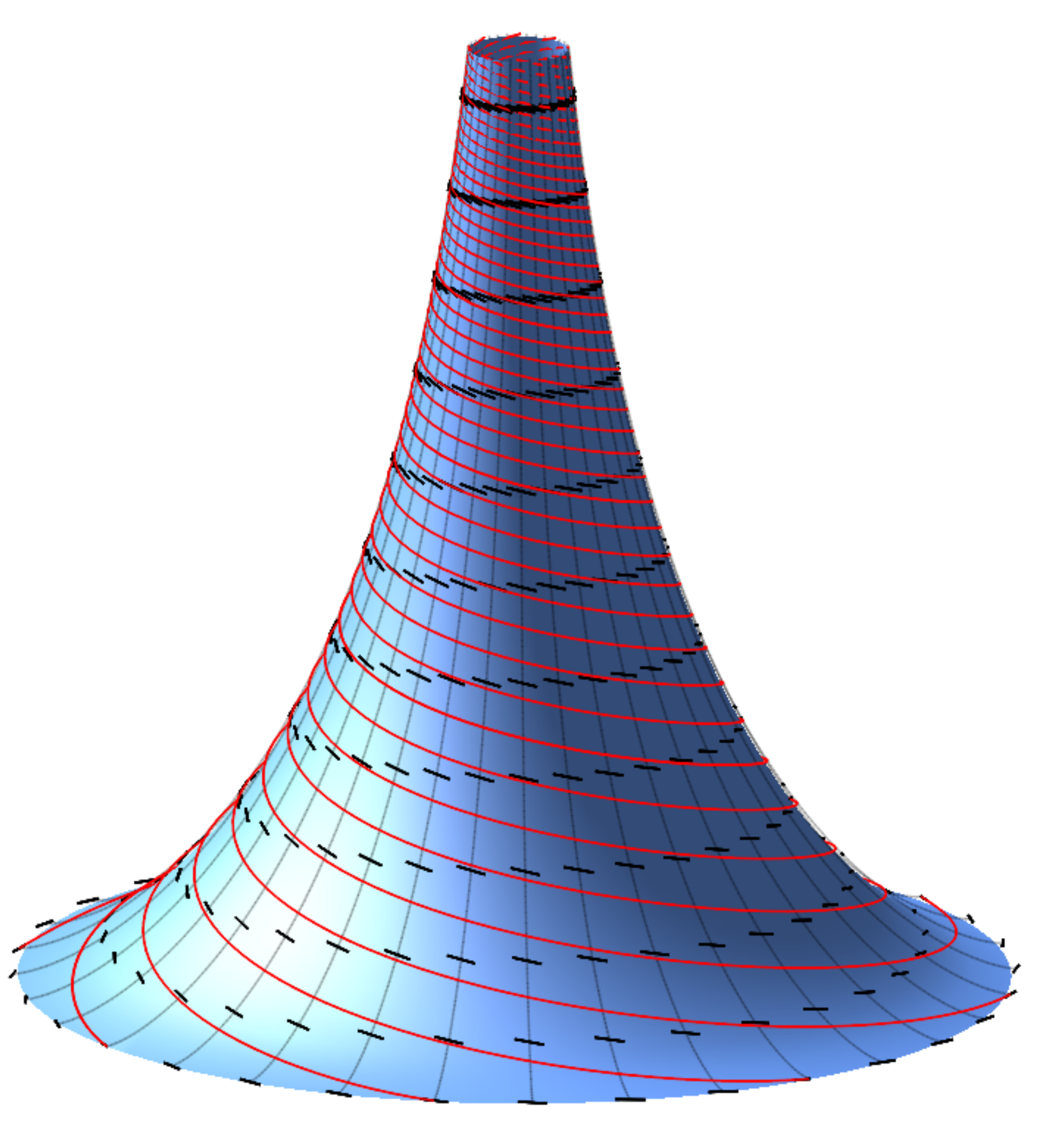}
	\caption{$\alpha=-\frac{\pi}{12}$}
	\label{fig:meridians_d}
\end{subfigure}
	\caption{Meridians are uniform geodesics of the pseudosphere. They convey a uniform field $\n$ (represented by headless directors) whose (red) field lines are loxodromes.}
	\label{fig:meridians}
\end{figure}

Figure~\ref{fig:meridians} illustrates different instances in which meridians convey a uniform field, with different choices of bend and splay reflected by different values of the angle $\alpha$ in \eqref{eq:n_n_perp_representations}.
The field lines of $\n$ (also shown in Fig.~\ref{fig:meridians}) are \emph{loxodromes} of the pseudosphere.\footnote{A loxodrome cuts meridians at a constant angle.}

As shown in Appendix~\ref{sec:ancillary}, a geodesic represented by \eqref{eq:pseudosphere_geodesic_coordinates} has unit tangent vector $\tangent$ on $\surface$  given by\footnote{Here $\tangent$ is oriented coherently with the positive orientation chosen for $\vt$ in Fig.~\ref{fig:pseudosphere_geodesics}.}
\begin{equation}
	\label{eq:pseudosphere_t}
	\tangent=-\sin\vt\e_u+\cos\vt\e_v,
\end{equation}
so that 
\begin{equation}
	\label{eq:pseudosphere_tperp}
	\tperp=-\cos\vt\e_u-\sin\vt\e_v.
\end{equation}
To apply \eqref{eq:uniform_geodesic_condition} to the general system of geodesics represented by \eqref{eq:pseudosphere_geodesic_coordinates}, we need to compute $\nablas\tangent$; it follows from \eqref{eq:pseudosphere_tperp} that 
\begin{equation}
	\label{eq:pseudosphere_nablas_t}
	\nablas\tangent=-\sin\vt\nablas\e_u+\cos\vt\nablas\e_v+\tperp\otimes\nablas\vt.
\end{equation}
Since (see Appendix~\ref{sec:ancillary}),
\begin{equation}
	\label{eq:pseudosphere_nablas_theta}
	\nablas\vt=\frac{\sin\vt}{u_0'\cos\vt+\rho'}\{-\rho'\sin\vt\e_u+(u_0'+\rho'\cos\vt)\e_v\},
\end{equation}
by \eqref{eq:pseudosphere:gliding_laws}, \eqref{eq:pseudosphere_tperp}, and \eqref{eq:pseudosphere_nablas_theta}, we easily arrive at
\begin{equation}
	\label{eq:pseudosphere_uniformity_condition}
	\tperp\cdot(\nablas\tangent)\tperp=-\frac{u_0'+\rho'\cos\vt}{u_0'\cos\vt+\rho'}.
\end{equation}
Thus, \eqref{eq:uniform_geodesic_condition} is satisfied if and only if
\begin{equation}
	\label{eq:pseudosphere_uniformity_equation}
	u_0'=\pm\rho',
\end{equation}
that is, whenever
\begin{equation}
	\label{eq:pseudosphere_uniformity_solution}
	u_0=\pm\rho+m.
\end{equation}

\begin{figure}[h] 
	\centering
	\begin{subfigure}[c]{0.3\linewidth}
		\centering
		\includegraphics[width=\linewidth]{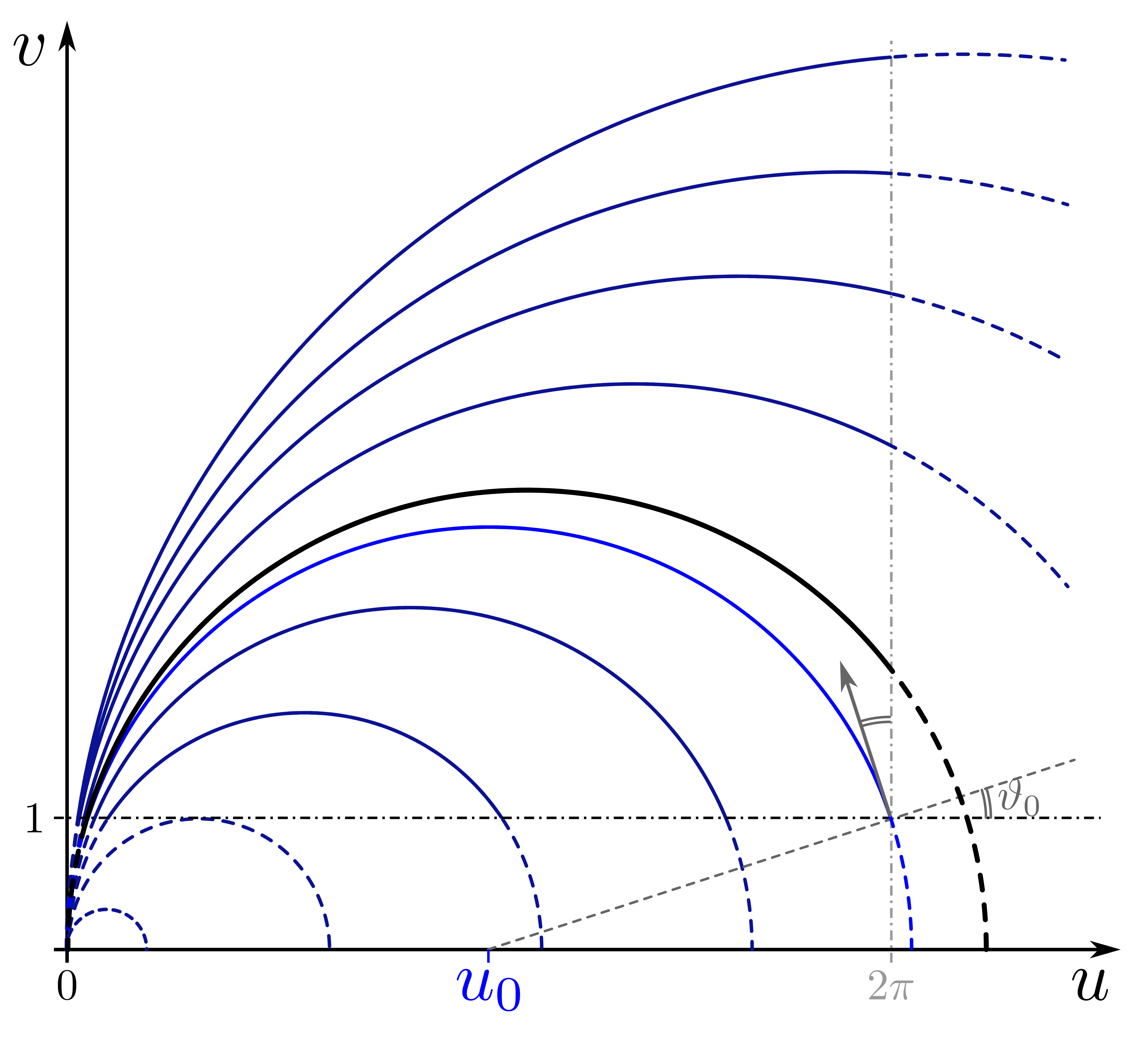}
		\caption{Right uniform geodesics.\\$m=0$}
		\label{fig:uniform_geodesics_a}
		\end{subfigure}
	\quad
	\begin{subfigure}[c]{0.3\linewidth}
	\centering
	\includegraphics[width=\linewidth]{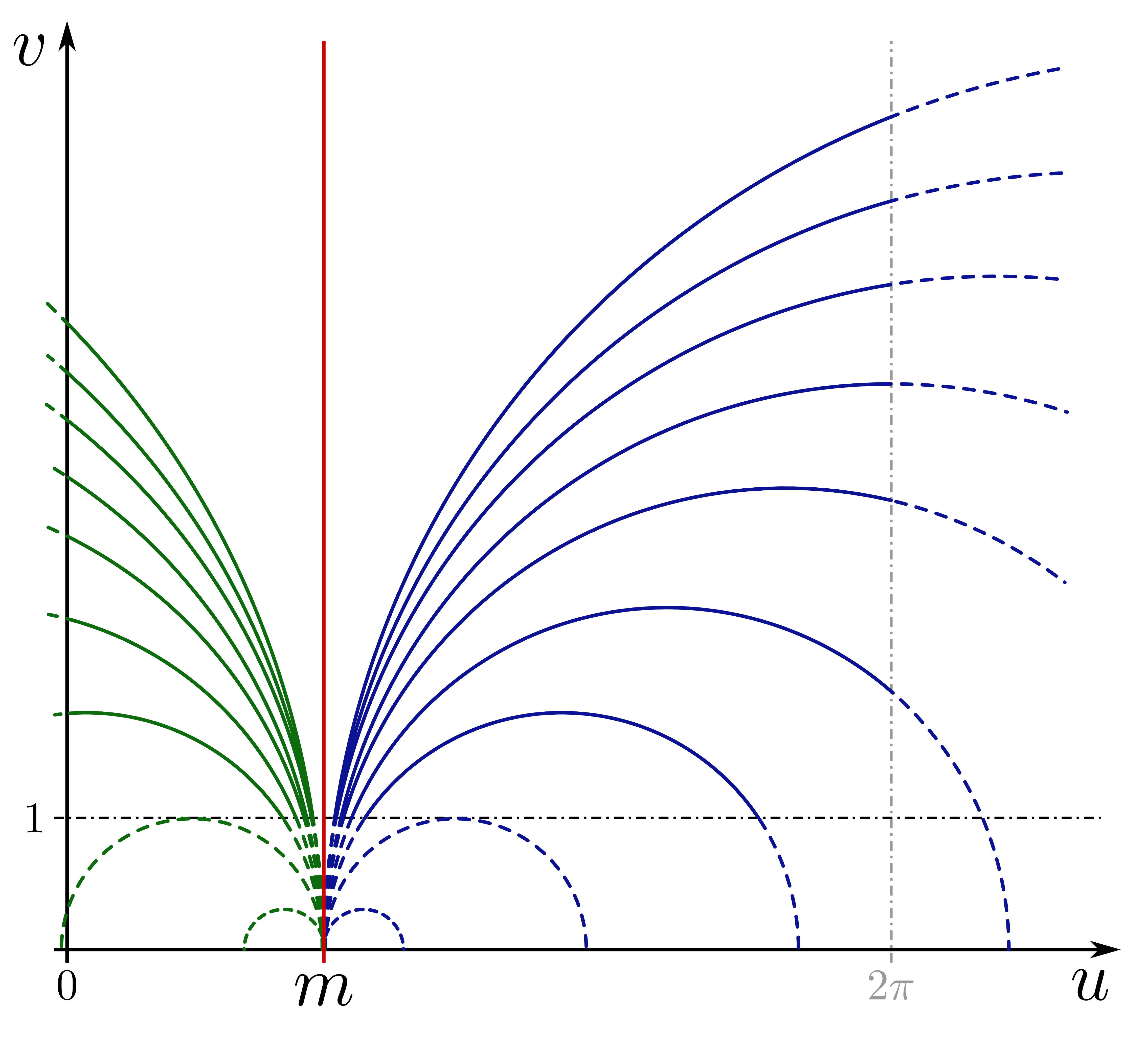}
	\caption{Right and left uniform geodesics.\\$0<m<2\pi$}
	\label{fig:uniform_geodesics_b}
\end{subfigure}
\quad 
\begin{subfigure}[c]{0.3\linewidth}
	\centering
	\includegraphics[width=\linewidth]{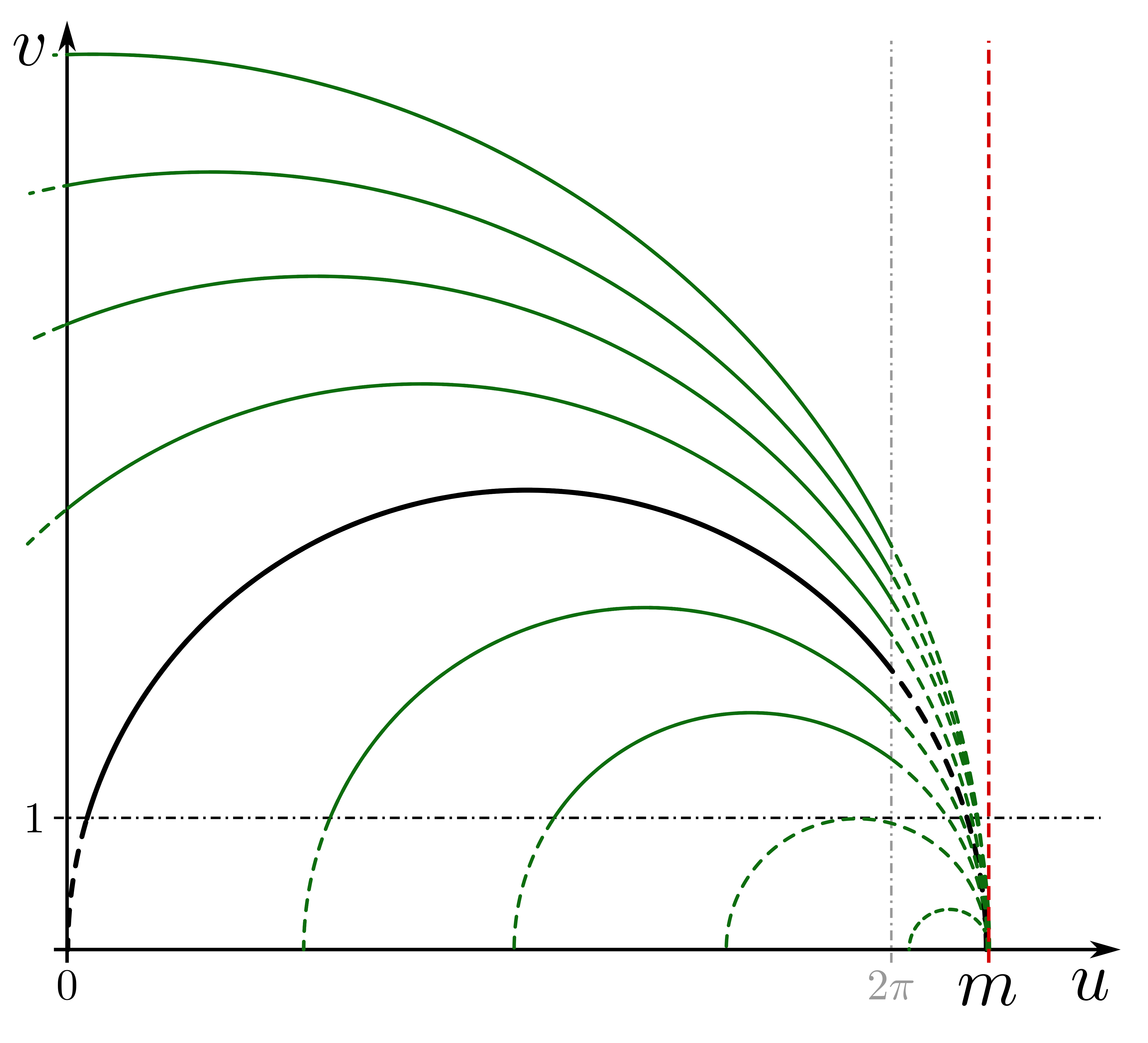}
	\caption{Left uniform geodesics.\\$m>2\pi$}
	\label{fig:uniform_geodesics_c}
\end{subfigure}
\caption{Pre-images of uniform geodesics shown in the $(u,v)$ plane for different values of the parameter $m$ designating the separating meridian (marked in red). Right uniform geodesics are blue, while left uniform geodesics are  green. The (exceptionally) black geodesic in panels (a) and (c) is precisely the same curve, belonging to two separate systems of right and left uniform geodesics.}
	\label{fig:uniform_geodesics}
\end{figure}

Geometrically, by \eqref{eq:pseudosphere_geodesic_coordinates}, \eqref{eq:pseudosphere_uniformity_solution} represents families of circles in the $(u,v)$ plane with centers on the $u$-axis, all tangent to one another at the point $(m,0)$, see Fig.~\ref{fig:uniform_geodesics}: for definiteness, we shall call \emph{right} and \emph{left} uniform geodesics those for which \eqref{eq:pseudosphere_uniformity_solution} holds with either the \emph{plus} or \emph{minus} sign, respectively.
The pre-images of uniform geodesics in Fig.~\ref{fig:uniform_geodesics} are drawn only within the domain $\Omega$, where the change of variables in \eqref{eq:pseudosphere_invertibility} is meaningful. Right and left uniform geodesics are separated by the meridian at $u=m$. 
Since the lines $u=0$ and $u=2\pi$ correspond to the same meridian on the pseudosphere, all uniform fields conveyed by these families of geodesics exhibit a \emph{line defect} along that (singular) meridian.

The reader should not be induced to think that a \emph{single} geodesic of $\surface$ could be designated as \emph{uniform}. As already apparent from \eqref{eq:uniform_geodesic_condition}, which involves $\nablas\tangent$ and not just $\tangent$, the notion of uniformity properly applies to a \emph{system} of geodesics, that is, to the way geodesics are bundled together. This is visually illustrated in Figs.~\ref{fig:uniform_geodesics_a} and \ref{fig:uniform_geodesics_c}: they show the same pre-image of a geodesic of $\surface$, seen  in one case as part of a system of right uniform geodesics and in the other of a system of left ones. Thus, if the direction of $\n$ is prescribed at a point on $\surface$ and an angle $\alpha$ is assigned, which represents constant distortion components, one would determine through \eqref{eq:t_choices} the direction of the tangent $\tangent$ to the local geodesic $\Curve$  that conveys  the uniform field  fulfilling these prescriptions. Away from $\Curve$,  however, there are \emph{two} uniform director fields with equal distortion components and the same trace on $\Curve$, one for each family of uniform geodesics (right and left) to which $\Curve$ belongs.

The dichotomy between right and left uniform geodesics on the pseudosphere is the visual embodiment of the duality between uniform fields anticipated on analytical grounds in Sec.~\ref{sec:parallel}. We may assert that such a duality actually persists on all pseudospherical surfaces.\footnote{Meridians are limiting cases of both right and left uniform geodesics. We may also say that they are self-conjugated, as they satisfy both forms of \eqref{eq:uniform_geodesic_condition}.}

\begin{figure}[h] 
	\centering
	\begin{subfigure}[c]{0.4\linewidth}
		\centering
		\includegraphics[width=\linewidth]{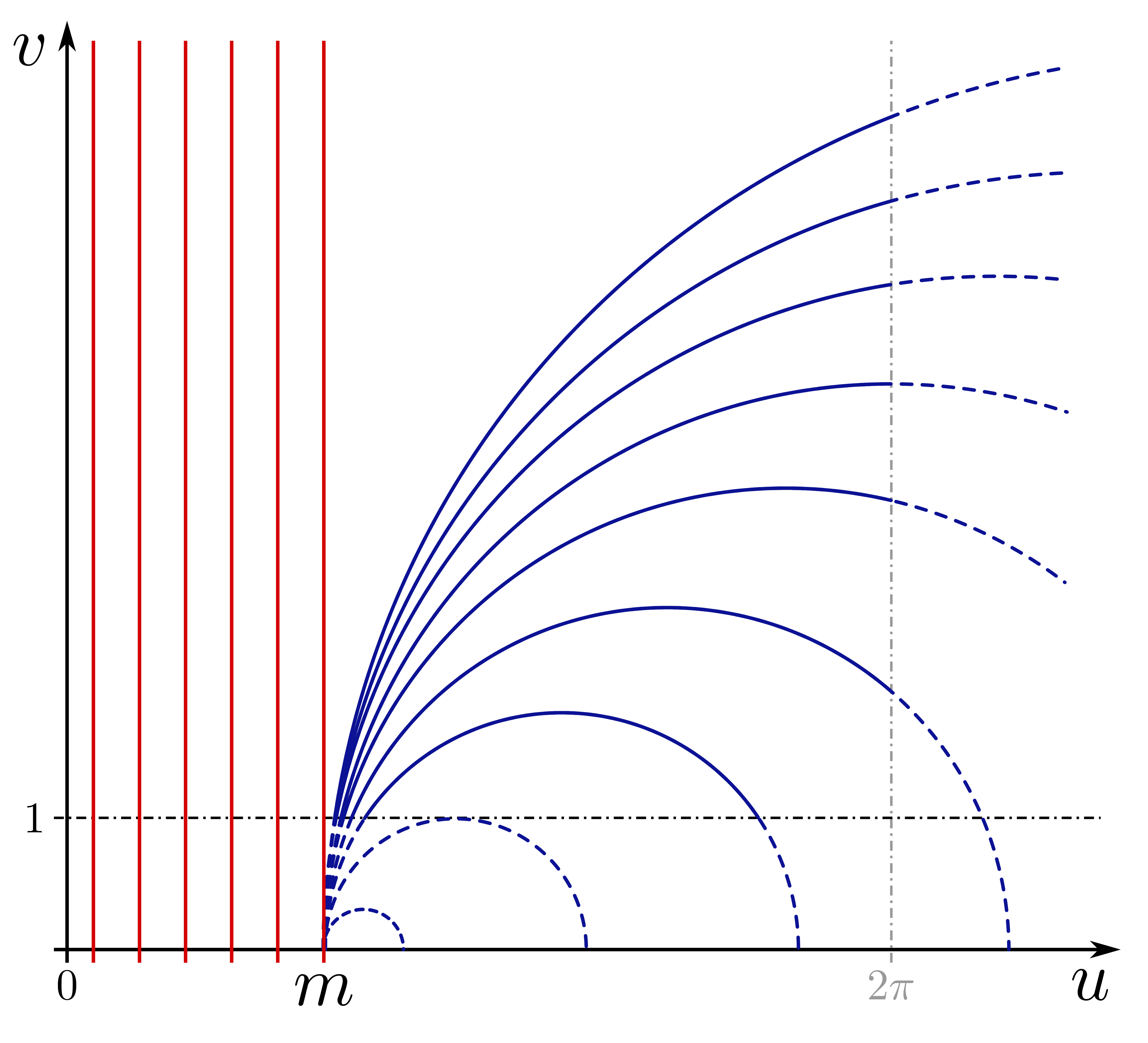}
		\caption{Meridians and right uniform.}
		\label{fig:hybridation_a}
	\end{subfigure}
	\quad
	\begin{subfigure}[c]{0.4\linewidth}
		\centering
		\includegraphics[width=\linewidth]{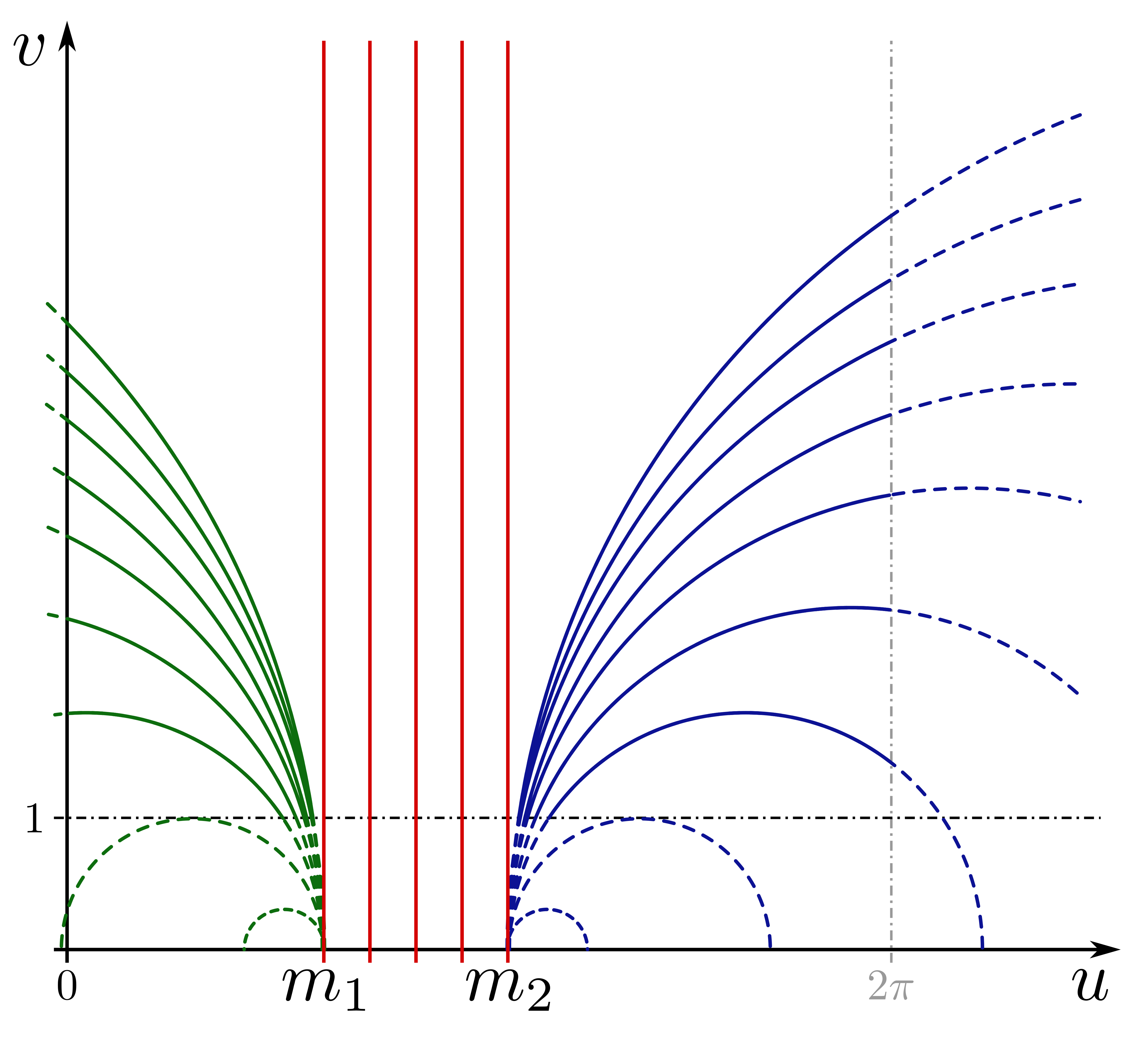}
		\caption{Meridians, right, and left uniform.}
		\label{fig:hybridation_b}
	\end{subfigure}
\caption{Hybrid uniform geodesics.}
\label{fig:hybridation}
\end{figure}

The separating meridian at $u=m$ in Fig.~\ref{fig:uniform_geodesics_b} is reached by both right and left uniform geodesics in the limit as $\rho\to+\infty$ (and, correspondingly, $u_0\to\pm\infty$) in \eqref{eq:pseudosphere_uniformity_solution}. Thus, by expanding the separating meridian, right and left uniform geodesics can be further combined, as shown by the examples in Fig.~\ref{fig:hybridation}.
These hybridations produce uniform fields that are of class $C^1$ (away from the singular meridian), but not $C^2$, due to the intervening separating meridians that can only be  approached asymptotically; they would be ruled out by a request of higher regularity.

\begin{figure}[h]
	\centering
	\includegraphics[width=.4\linewidth]{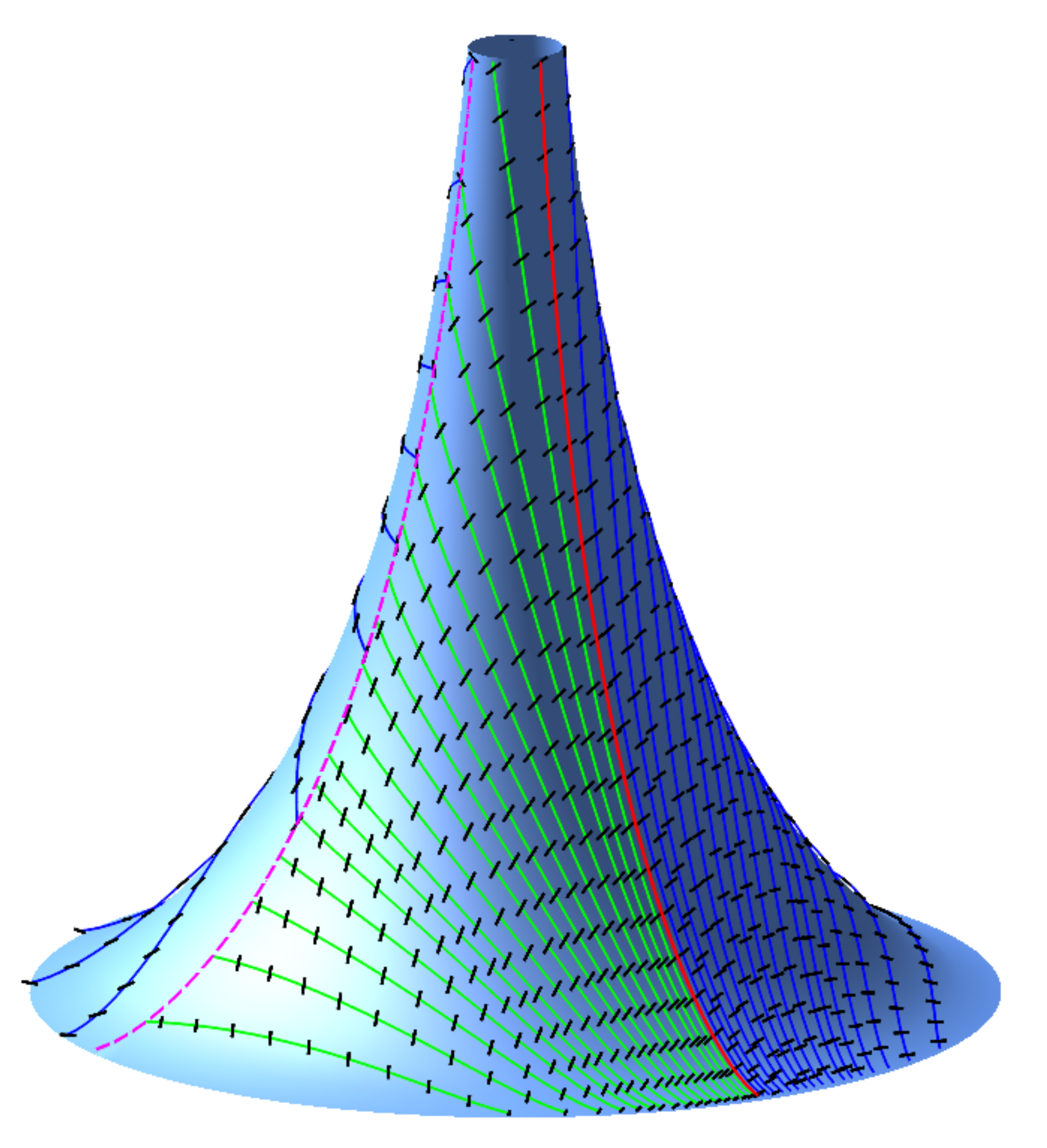}
\caption{Uniform geodesics as in Fig.~\ref{fig:uniform_geodesics_b} conveying a uniform field $\n$ as in \eqref{eq:n_n_perp_representations} with $\alpha=-\pi/3$. The separating meridian (at $u=m$) is red, while the singular meridian (at $u=0$) is (dashed) purple. Right uniform geodesics are blue, while left uniform geodesics are green.}
\label{fig:ostracism}
\end{figure}

The same ostracism would fall on the case shown in Fig.~\ref{fig:uniform_geodesics_b}, for which we illustrate in Fig.~\ref{fig:ostracism} the corresponding uniform geodesics conveying a uniform director field (represented by headless directors) with $\alpha=-\pi/3$.

\begin{figure}[h] 
	\centering
	\begin{subfigure}[c]{0.3\linewidth}
		\centering
		\includegraphics[width=\linewidth]{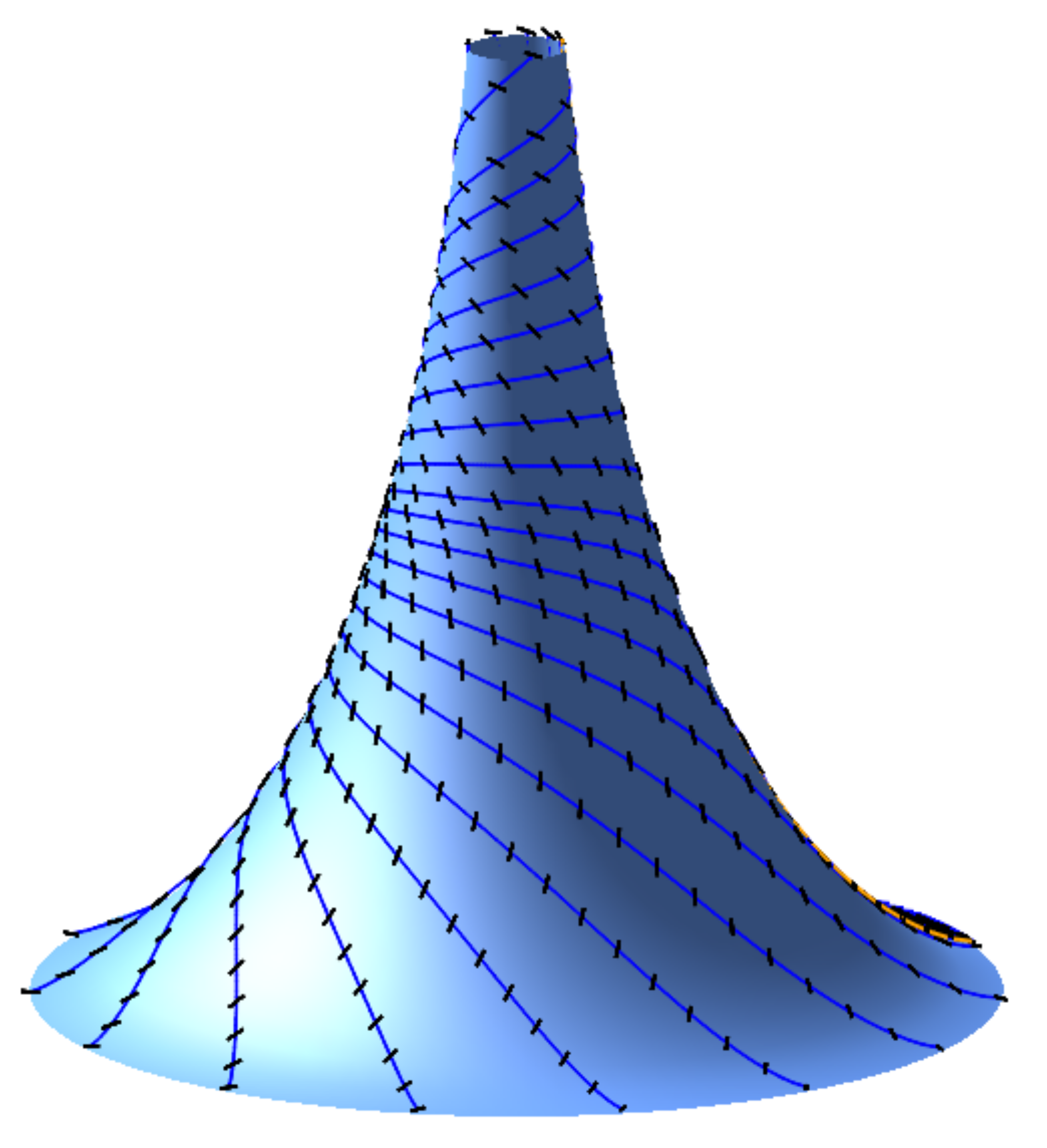}
		\caption{}
		\label{fig:galley_a}
	\end{subfigure}
	\quad
	\begin{subfigure}[c]{0.3\linewidth}
		\centering
		\includegraphics[width=\linewidth]{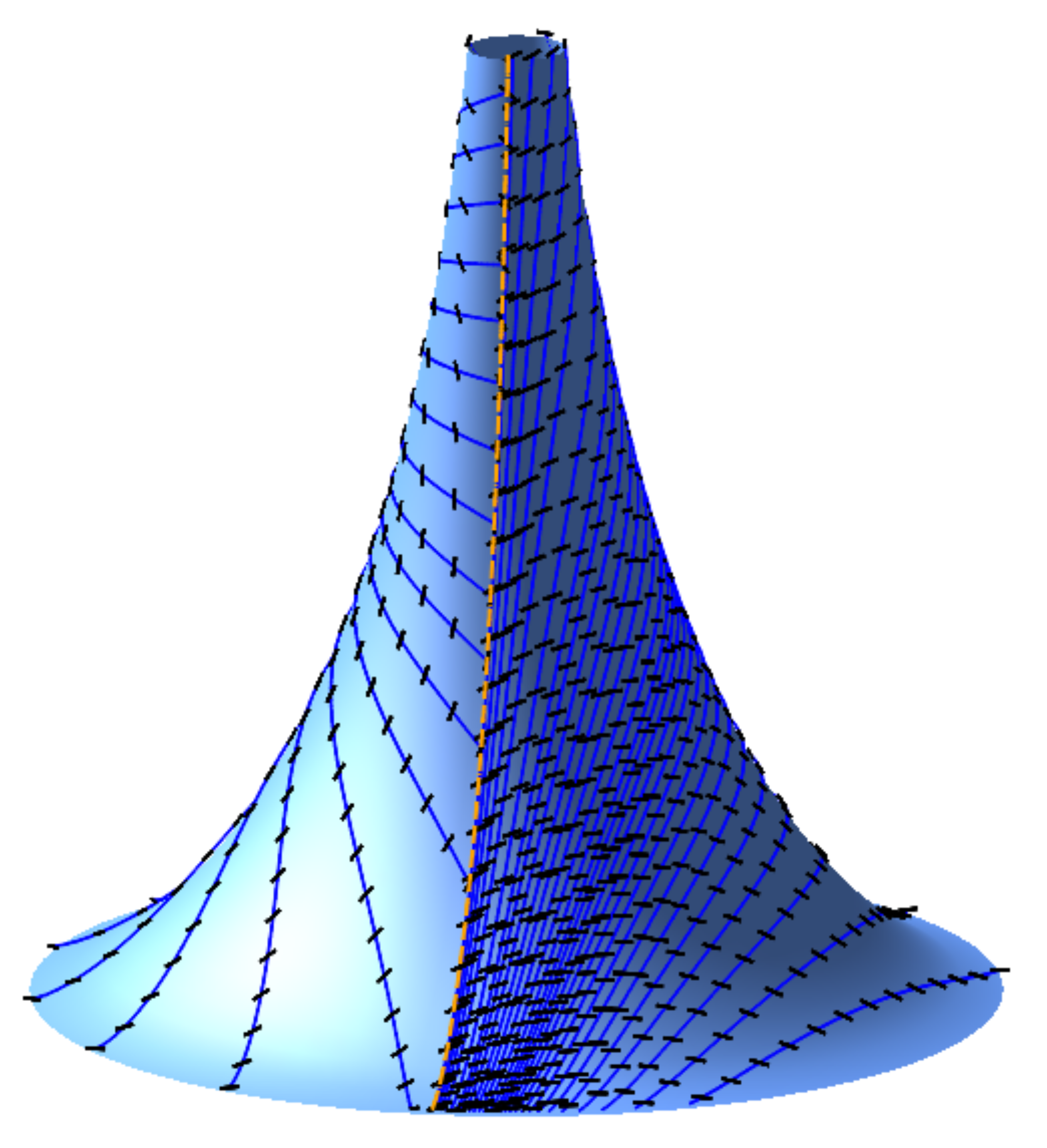}
		\caption{}
		\label{fig:gallery_b}
	\end{subfigure}
\quad
\begin{subfigure}[c]{0.3\linewidth}
	\centering
	\includegraphics[width=\linewidth]{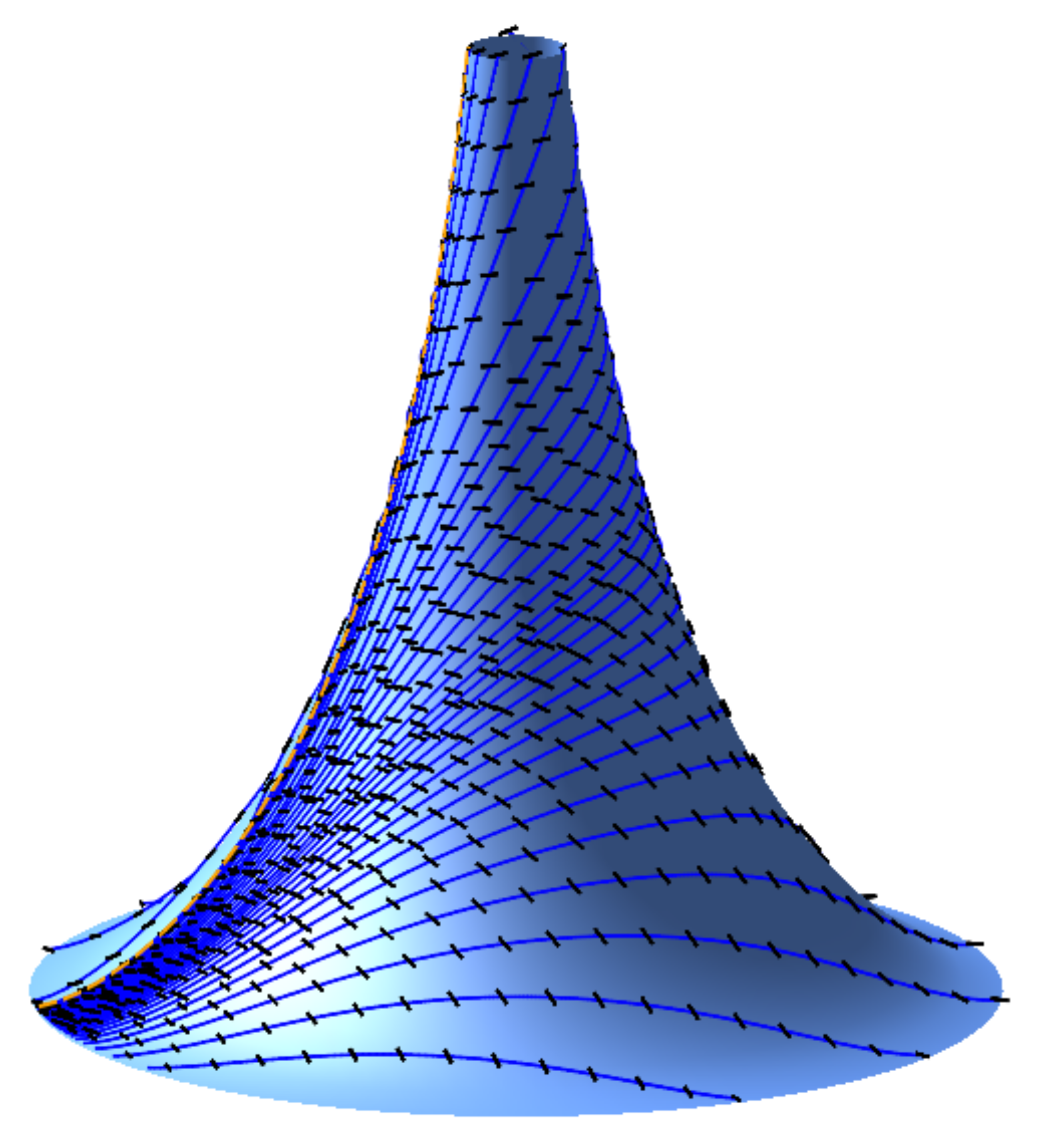}
	\caption{}
	\label{fig:gallery_c}
\end{subfigure}
\\
\begin{subfigure}[c]{0.3\linewidth}
	\centering
	\includegraphics[width=\linewidth]{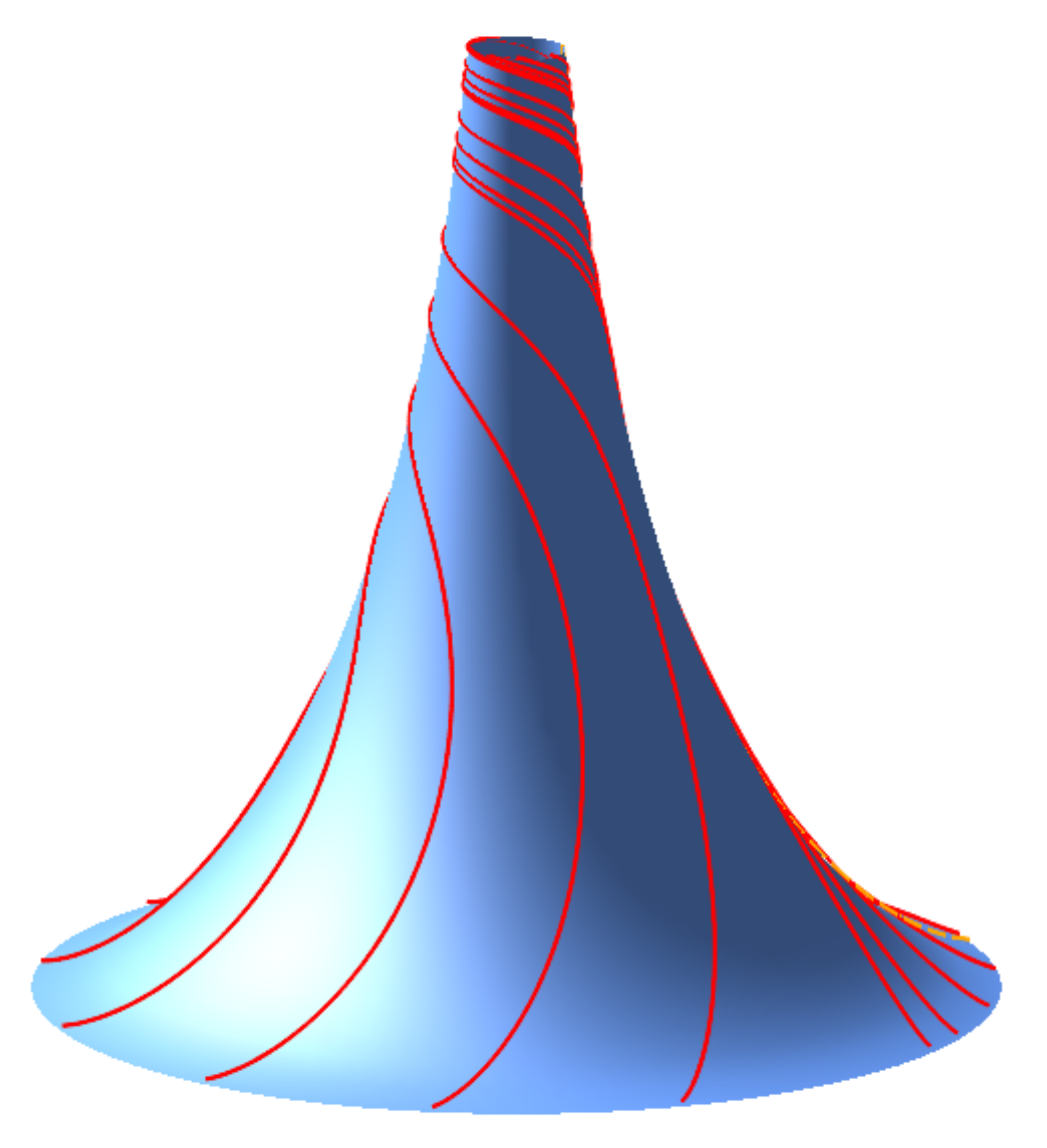}
	\caption{}
	\label{fig:galley_d}
\end{subfigure}
\quad
\begin{subfigure}[c]{0.3\linewidth}
	\centering
	\includegraphics[width=\linewidth]{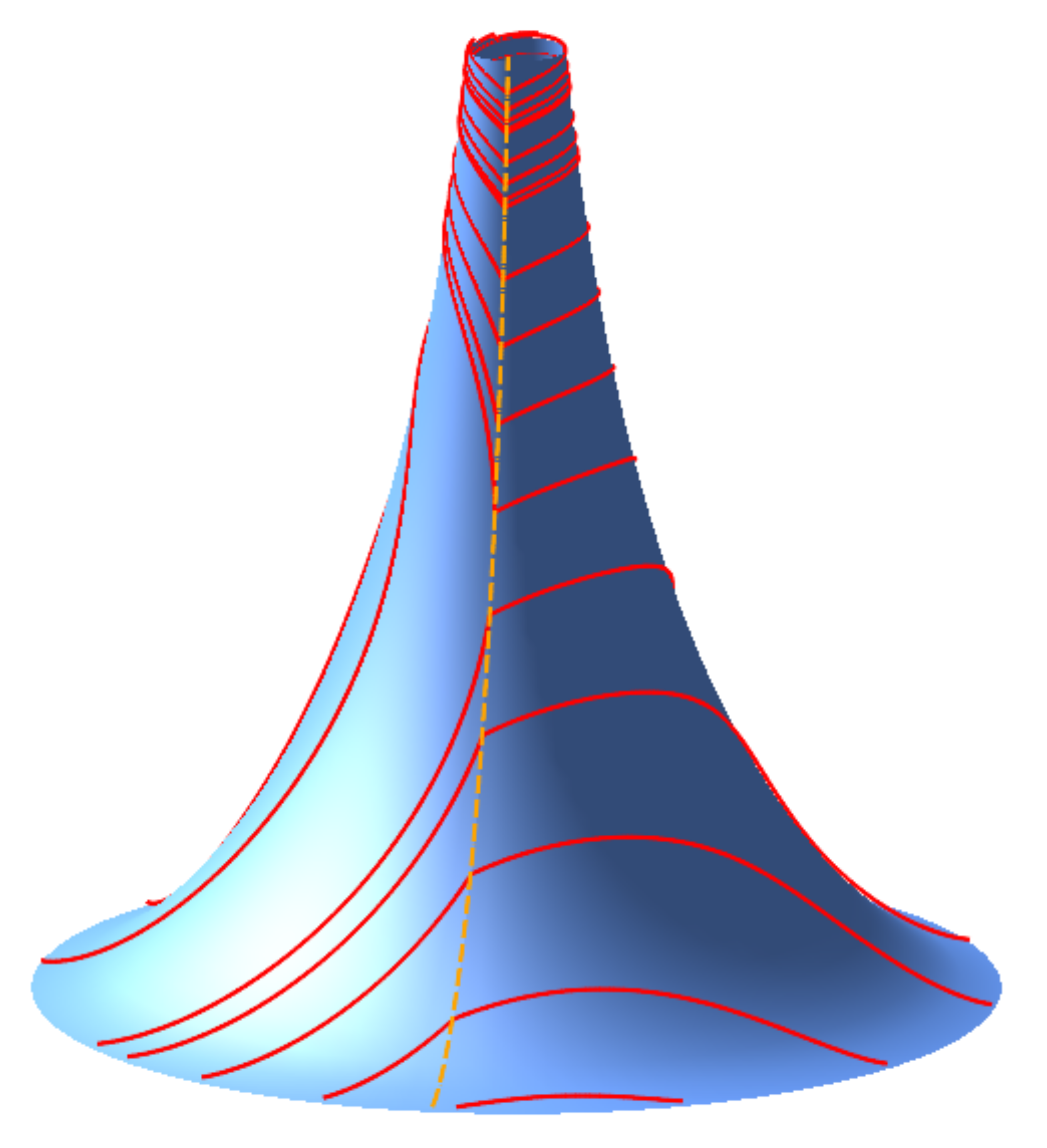}
	\caption{}
	\label{fig:gallery_e}
\end{subfigure}
\quad
\begin{subfigure}[c]{0.3\linewidth}
	\centering
	\includegraphics[width=\linewidth]{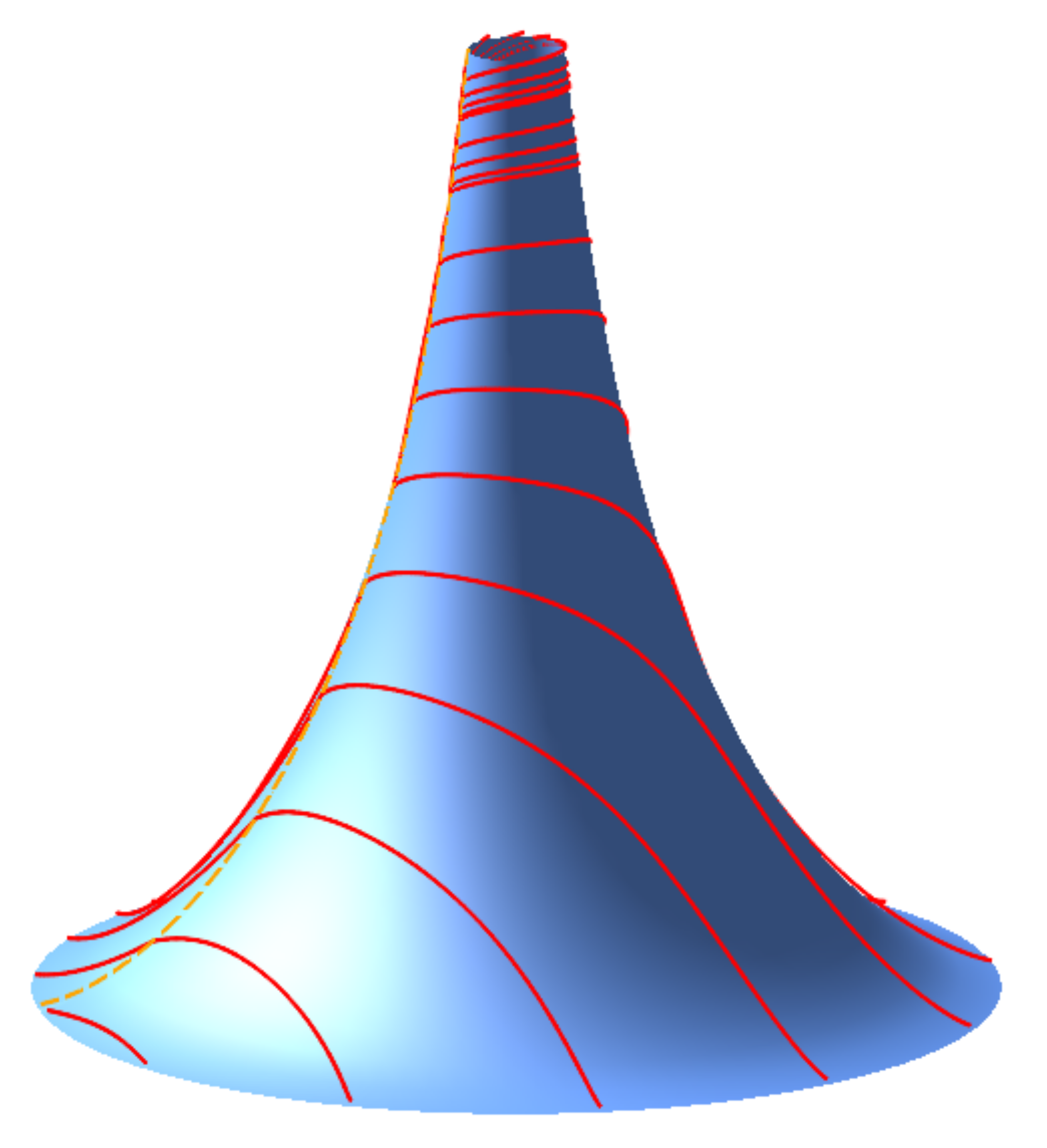}
	\caption{}
	\label{fig:gallery_f}
\end{subfigure}
	\caption{Three views of the pseudosphere bearing the uniform director field $\n$ in \eqref{eq:n_n_perp_representations} with $\alpha=-\pi/3$. The pre-images of the conveying geodesics are depicted in Fig.~\ref{fig:uniform_geodesics_a}. Geodesics and directors are shown in panels (a), (b), and (c), while the corresponding field lines of $\n$ are shown in panels (d), (e), and (f). Geodesics are blue, nematic field lines are red, and the singular meridian is yellow. The supplementary movie shows all other views.}
	\label{fig:gallery}
\end{figure}

To illustrate a case of higher regularity, we set $m=0$, as in Fig.~\ref{fig:uniform_geodesics_a}, and draw via \eqref{eq:pseudosphere:r} the corresponding family of (right) uniform geodesics on the pseudosphere. Both uniform geodesics and director fields for selected distortion components (again $\alpha=-\pi/3$) are depicted in Fig.~\ref{fig:gallery}, where three views are shown; an animation where the pseudosphere rotates about its symmetry axis is provided as a supplementary material accompanying this paper.

For $m=0$, as a measure of defectiveness on the singular meridian, we compute the maximum angular mismatch of directors along it. As is easily seen, this is independent of $\alpha$, it is attained on the bounding rim, and equals the angle $\vt_0$ shown in Fig.~\ref{fig:uniform_geodesics_a}; a simple geometric construction delivers
\begin{equation}
	\label{eq:pseudosphere_theta_0}
	\vt_0=\arccos\left(\frac{4\pi^2-1}{4\pi^2+1}\right).
\end{equation}

By choosing $m<0$ (or $m>2\pi$, for left uniform geodesics), other, less distorted uniform fields can easily be generated. In the limit as $m\to-\infty$ (or $m\to+\infty$, for left uniform geodesics), these recover asymptotically the meridians of the pseudosphere.

\section{Conclusions}\label{sec:conclusions}
We have addressed the problem of determining the most general nematic field on a smooth surface that a two-dimensional observer (intrinsic to the surface and unaware of the dimension along the normal) would see as equally distorted  at all points. For these fields, called uniform, we gave a definition equivalent to the standard one, but formulated in the alternative language of moving frames.

Nematic uniformity touches upon the frustrating power of surfaces. The intuitive idea behind this association is that the ground state of any elastic theory based on a surface director order parameter should possibly be uniform. Were the latter impeded, it would ignite geometric frustration.

A remarkable advance in the study of surface uniformity was the necessary condition proved by Niv and Efrati \cite{niv:geometric} that requires the hosting surface to be pseudospherical, that is, with constant \emph{negative} Gaussian curvature. However, nothing was known about the actual existence of uniform fields on those surfaces, let alone their structure.

We solved this problem
\begin{inparaenum}[(1)]
	\item by proving that a uniform field is parallel transported by geodesics and
	\item by characterizing all systems of geodesics that can convey a uniform field.
\end{inparaenum}

The latter were said to constitute a system of uniform geodesics, reflecting more the way they are bundled together than an individual property. For any given geodesic, we found two distinct \emph{systems} of uniform geodesics to which it belongs, all conveying a uniform field. We conventionally called \emph{right} and \emph{left} these systems, thus alluding at a possible intrinsic way to introduce handedness on a surface.

Our general geometric construction was made explicit for Beltrami's pseudosphere, whose uniform fields were characterized completely. These are both without and with defects. The former have the loxodromes of meridians as field lines, all others have a line defect along a singular meridian.

Since, by Minding's classical theorem, all surfaces with the same constant Gaussian curvature are isometric and both geodesics and uniformity are preserved by isometries, the solution for the pseudosphere entrains the solution for all pseudospherical surfaces: it would suffice to carry over (at least locally) the system of uniform geodesics.

Although this last task may fail to be easily accomplished, the general structure of uniform geodesics and the corresponding duality between the generated uniform fields remain valid. An analytic condition was given that characterized the system of uniform geodesics on a generic surface. Solving it, one can find directly all uniform fields on that surface.

Our study was confined to smooth surfaces imbedded in three-dimensional space. We wonder whether systems of uniform geodesics would exist in more general differential manifolds.

Even for a surface, we are intrigued by the conjugation by duality found among uniform nematic fields. We wonder whether this could introduce another intrinsic notion of planar \emph{chirality} (see, for example, \cite{bohmer:chirality} for an account on the existing ones).

\begin{acknowledgements}
Both authors are members of \emph{GNFM}, a branch of \emph{INdAM}, the Italian Institute for Advanced Mathematics. A.P. wishes to acknowledge financial support from the Italian MIUR through European Programmes REACT EU 2014-2020 and PON 2014-2020 CCI2014IT16M2OP005.
\end{acknowledgements}

\appendix
\section{Analytic Characterization of Uniformity }\label{sec:conditions}
This appendix collects conditions for surface uniformity that involve the connectors of a moving frame. They are necessary and sufficient, but hard to resolve; they are recorded here for possible later use, but mainly to provide an alternative, independent proof of the known fact that a uniform field can only be exhibited by a surface with constant negative Gaussian curvature.

In tune with the proof of isometric invariance of uniformity presented in Sect.~\ref{sec:uniform},  here we employ systematically the (necessary and sufficient) integrability  condition in \eqref{eq:integrability_vector} to characterize the connectors of the moving frames compatible with a uniform unit vector field $\n$. In accordance with \eqref{eq:c_1_c_2}, we start by writing $\cv$ in \eqref{eq:gliding_laws_n} as
\begin{equation}
	\label{eq:c_uniform}
	\cv=-b_\perp\n+S\nperp.
\end{equation}
To apply \eqref{eq:integrability_vector} to $\H=\nablas\n$, we first insert \eqref{eq:c_uniform} in \eqref{eq:gliding_laws_n} and, since both $b_\perp$ and $S$ are constant, we obtain that
\begin{subequations}\label{eq:integrability_a_b}
\begin{equation}
	\label{eq:integrability_a}
	\begin{split}
	2\skw(\nablastwo\n)&=\nperp\otimes[(b_\perp^2+S^2)(\nperp\otimes\n-\n\otimes\nperp)+b_\perp(\dv_1\otimes\normal-\normal\otimes\dv_1)\\&\qquad\qquad-S(\dv_2\otimes\normal-\normal\otimes\dv_2)+(\dv_2\otimes\dv_1-\dv_1\otimes\dv_2)]\\
	&+\normal\otimes[b_\perp(\dv_2\otimes\n-\n\otimes\dv_2)-S(\dv_2\otimes\nperp-\nperp\otimes\dv_2)+(\nablas\dv_1)-(\nablas\dv_1)\trans],
	\end{split}
\end{equation}
\begin{equation}
	\label{eq:integrability_b}
	\begin{split}
		2\skw[(\nablas\n)\curvature\otimes\normal]&=\nperp\otimes[b_\perp(\dv_1\otimes\normal-\normal\otimes\dv_1)-S(\dv_2\otimes\normal-\normal\otimes\dv_2)]\\
		&-\normal\otimes[(\dv_1\cdot\n)(\dv_1\otimes\normal-\normal\otimes\dv_1)+(\dv_2\cdot\n)(\dv_2\otimes\normal-\normal\otimes\dv_2)],
	\end{split}
\end{equation}
\end{subequations}
where use of \eqref{eq:connector_identity_n} has also been made. 

We then require that the right-hand sides of equations \eqref{eq:integrability_a_b} be equal to one another. This amounts to two equalities between axial vectors, one for each non-vanishing left component of the third-rank tensors on \eqref{eq:integrability_a_b},
\begin{equation}
	\label{eq:pre_Gaussian}
	(b_\perp^2+S^2)\normal+\dv_1\times\dv_2=\zero,
\end{equation}
\begin{equation}
	\label{eq:d_1_pde}
	\curls\dv_1=(\cv_\perp\cdot\dv_2)\normal-(\dv_1\otimes\dv_1+\dv_2\otimes\dv_2)\nperp,
\end{equation}
where $\cv_\perp:=\normal\times\cv=-S\n-b_\perp\nperp$. It readily follows from \eqref{eq:gaussian_curvature} that \eqref{eq:pre_Gaussian} can be written in the equivalent form,
\begin{equation}
	\label{eq:Gaussian_curvature_condition}
	b_\perp^2+S^2=-K,
\end{equation}
which requires $\surface$ to be a surface with constant \emph{negative} Gaussian curvature related to the bend and splay components of the uniform field $\n$, as already proved by a different method in \cite{niv:geometric}.

By applying the same integrability condition to $\H=\nablas\normal$, we obtain again \eqref{eq:Gaussian_curvature_condition}, accompanied this time by another equation, which supplements \eqref{eq:d_1_pde},
\begin{equation}
	\label{eq:d_2_pde}
	\curls\dv_2=-(\cv_\perp\cdot\dv_1)\normal+(\dv_1\otimes\dv_1+\dv_2\otimes\dv_2)\n.
\end{equation}
It is not necessary to require explicitly the integrability of $\nablas\n_\perp$, as this is already implicit in \eqref{eq:d_1_pde}, \eqref{eq:Gaussian_curvature_condition}, and \eqref{eq:d_2_pde}, once we set $\nperp=\normal\times\n$.

Equations \eqref{eq:d_1_pde}, \eqref{eq:Gaussian_curvature_condition}, and \eqref{eq:d_2_pde} constitute a system of necessary and sufficient conditions for the existence of a surface uniform field phrased in terms of the connectors of a moving frame. Of course, taken by itself, \eqref{eq:Gaussian_curvature_condition} is only necessary: it prescribes the class of background surfaces upon which the first-order partial differential equations \eqref{eq:d_1_pde} and \eqref{eq:d_2_pde} should be integrated. The latter is not an easy task: both \eqref{eq:d_1_pde} and \eqref{eq:d_2_pde} can indeed be made to depend only on the fields $\dv_1$, $\dv_2$ (and, of course, $\normal$), by use of the equations,
\begin{equation}\label{eq:n_n_perp_d_1_d_2}
	\n=\curvature^{-1}\dv_1,\qquad\nperp=\curvature^{-1}\dv_2,
\end{equation}
as, by \eqref{eq:Gaussian_curvature_condition}, the curvature tensor $\nablas\normal$ is invertible on the local tangent plane.\footnote{It readily follows from \eqref{eq:n_n_perp_d_1_d_2} that the identity \eqref{eq:connector_identity_n} is automatically satisfied.} Once the fields $\dv_1$ and $\dv_2$ that solve \eqref{eq:d_1_pde} and \eqref{eq:d_2_pde} are known, the first equation in \eqref{eq:n_n_perp_d_1_d_2} then delivers the desired uniform field $\n$.

The method illustrated in the main text to find all surface uniform fields has luckily followed a different, more geometric avenue.

\section{Gradient in Geodesic Coordinates}\label{sec:ancillary}
In this Appendix, we learn how to express the surface gradient on the pseudosphere $\surface$ of a differentiable function $\psi(\vp,\vt)$ that depends on the geodesic coordinates $(\vp,\vt)$ defined in \eqref{eq:pseudosphere_geodesic_coordinates}.

We start by considering a curve $t\mapsto\bm{p}(t)$ generated on $\surface$ by taking $\vp=\vp(t)$ and $\vt=\vt(t)$,
\begin{equation}
	\label{eq:curve}
	\bm{p}(t)=\rv(u(t),v(t)),
\end{equation}
where the functions $u(t)$ and $v(t)$ are obtained by composing $(\vp(t),\vt(t))$ with \eqref{eq:pseudosphere_geodesic_coordinates}. By the chain rule and use of \eqref{eq:e_u_e_v_definition}, since for $\rv$ in \eqref{eq:pseudosphere:r}
\begin{equation}
	\label{eq:r_u=r_v}
	|\rv_u|=|\rv_v|=\frac{1}{v},
\end{equation}
we readily see that 
\begin{equation}
	\label{eq:curve_dot}
	\dot{\bm{p}}=\frac{1}{\rho\sin\vt}\left\{[(u_0'+\rho'\cos\vt)\dot{\vp}-\rho\sin\vt\dot{\vt}]\e_u+(\rho'\sin\vt\dot{\vp}+\rho\cos\vt\dot{\vt})\e_v\right\},
\end{equation}
where a superimposed dot denotes differentiation with respect to the parameter $t$. By letting $\dot{\vp}=0$ in \eqref{eq:curve_dot}, we obtain \eqref{eq:pseudosphere_t} in the main text.

Moreover, by inserting \eqref{eq:curve_dot} into the identity
\begin{equation}
	\label{eq:gradient_identity}
	\dot{\psi}=\frac{\partial\psi}{\partial\vp}\dot{\vp}+\frac{\partial\psi}{\partial\vt}\dot{\vt}=\nablas\psi\cdot\dot{\bm{p}},
\end{equation}
by the arbitrarity of $\dot{\vp}$ and $\dot{\vt}$, we derive the following equations
\begin{subequations}
	\label{eq:gradient_psi_equations}
	\begin{numcases} {}
		\frac{1}{\rho\sin\vt}(u_0'+\rho'\cos\vt)\psi_u+\frac{\rho'}{\rho}\psi_v=\frac{\partial\psi}{\partial\vp},\\
	-\psi_u+\cot\vt\psi_v=\frac{\partial\psi}{\partial\vt},
	\end{numcases}
\end{subequations}
for the components $(\psi_u,\psi_v)$ of $\nablas\psi$ in the frame $(\e_u,\e_v)$. The solution of the linear system in \eqref{eq:gradient_psi_equations} is given by
\begin{subequations}
	\label{eq:psi_gradient_solution}
	\begin{numcases} {}
	\psi_u=\frac{\sin\vt}{u_0'\cos\vt+\rho'}\left(\rho\cos\vt\frac{\partial\psi}{\partial\vp}-\rho'\sin\vt\frac{\partial\psi}{\partial\vt}\right),\\
		\psi_v=\frac{\sin\vt}{u_0'\cos\vt+\rho'}\left(\rho\sin\vt\frac{\partial\psi}{\partial\vp}+(u_0'+\rho'\cos\vt)\frac{\partial\psi}{\partial\vt}\right),
	\end{numcases}
\end{subequations}
which for $\psi\equiv\vt$ deliver \eqref{eq:pseudosphere_nablas_theta} in the main text.


\begin{thebibliography}{37}%
	\makeatletter
	\providecommand \@ifxundefined [1]{%
		\@ifx{#1\undefined}
	}%
	\providecommand \@ifnum [1]{%
		\ifnum #1\expandafter \@firstoftwo
		\else \expandafter \@secondoftwo
		\fi
	}%
	\providecommand \@ifx [1]{%
		\ifx #1\expandafter \@firstoftwo
		\else \expandafter \@secondoftwo
		\fi
	}%
	\providecommand \natexlab [1]{#1}%
	\providecommand \enquote  [1]{``#1''}%
	\providecommand \bibnamefont  [1]{#1}%
	\providecommand \bibfnamefont [1]{#1}%
	\providecommand \citenamefont [1]{#1}%
	\providecommand \href@noop [0]{\@secondoftwo}%
	\providecommand \href [0]{\begingroup \@sanitize@url \@href}%
	\providecommand \@href[1]{\@@startlink{#1}\@@href}%
	\providecommand \@@href[1]{\endgroup#1\@@endlink}%
	\providecommand \@sanitize@url [0]{\catcode `\\12\catcode `\$12\catcode
		`\&12\catcode `\#12\catcode `\^12\catcode `\_12\catcode `\%12\relax}%
	\providecommand \@@startlink[1]{}%
	\providecommand \@@endlink[0]{}%
	\providecommand \url  [0]{\begingroup\@sanitize@url \@url }%
	\providecommand \@url [1]{\endgroup\@href {#1}{\urlprefix }}%
	\providecommand \urlprefix  [0]{URL }%
	\providecommand \Eprint [0]{\href }%
	\providecommand \doibase [0]{https://doi.org/}%
	\providecommand \selectlanguage [0]{\@gobble}%
	\providecommand \bibinfo  [0]{\@secondoftwo}%
	\providecommand \bibfield  [0]{\@secondoftwo}%
	\providecommand \translation [1]{[#1]}%
	\providecommand \BibitemOpen [0]{}%
	\providecommand \bibitemStop [0]{}%
	\providecommand \bibitemNoStop [0]{.\EOS\space}%
	\providecommand \EOS [0]{\spacefactor3000\relax}%
	\providecommand \BibitemShut  [1]{\csname bibitem#1\endcsname}%
	\let\auto@bib@innerbib\@empty
	\bibitem [{\citenamefont {Virga}(2019)}]{virga:uniform}%
	\BibitemOpen
	\bibfield  {author} {\bibinfo {author} {\bibfnamefont {E.~G.}\ \bibnamefont
			{Virga}},\ }\bibfield  {title} {\bibinfo {title} {Uniform distortions and
			generalized elasticity of liquid crystals},\ }\href
	{https://doi.org/10.1103/PhysRevE.100.052701} {\bibfield  {journal} {\bibinfo
			{journal} {Phys. Rev. E}\ }\textbf {\bibinfo {volume} {100}},\ \bibinfo
		{pages} {052701} (\bibinfo {year} {2019})}\BibitemShut {NoStop}%
	\bibitem [{\citenamefont {Selinger}(2018)}]{selinger:interpretation}%
	\BibitemOpen
	\bibfield  {author} {\bibinfo {author} {\bibfnamefont {J.~V.}\ \bibnamefont
			{Selinger}},\ }\bibfield  {title} {\bibinfo {title} {Interpretation of
			saddle-splay and the {O}seen-{F}rank free energy in liquid crystals},\ }\href
	{https://doi.org/https://doi.org/10.1080/21680396.2019.1581103} {\bibfield
		{journal} {\bibinfo  {journal} {Liq. Cryst. Rev.}\ }\textbf {\bibinfo
			{volume} {6}},\ \bibinfo {pages} {129} (\bibinfo {year} {2018})}\BibitemShut
	{NoStop}%
	\bibitem [{\citenamefont {Machon}\ and\ \citenamefont
		{Alexander}(2016)}]{machon:umbilic}%
	\BibitemOpen
	\bibfield  {author} {\bibinfo {author} {\bibfnamefont {T.}~\bibnamefont
			{Machon}}\ and\ \bibinfo {author} {\bibfnamefont {G.~P.}\ \bibnamefont
			{Alexander}},\ }\bibfield  {title} {\bibinfo {title} {Umbilic lines in
			orientational order},\ }\href
	{https://doi.org/https://doi.org/10.1103/PhysRevX.6.011033} {\bibfield
		{journal} {\bibinfo  {journal} {Phys. Rev. X}\ }\textbf {\bibinfo {volume}
			{6}},\ \bibinfo {pages} {011033} (\bibinfo {year} {2016})}\BibitemShut
	{NoStop}%
	\bibitem [{\citenamefont {Selinger}(2022)}]{selinger:director}%
	\BibitemOpen
	\bibfield  {author} {\bibinfo {author} {\bibfnamefont {J.~V.}\ \bibnamefont
			{Selinger}},\ }\bibfield  {title} {\bibinfo {title} {Director deformations,
			geometric frustration, and modulated phases in liquid crystals},\ }\href
	{https://doi.org/10.1146/annurev-conmatphys-031620-105712} {\bibfield
		{journal} {\bibinfo  {journal} {Ann. Rev. Condens. Matter Phys.}\ }\textbf
		{\bibinfo {volume} {13}},\ \bibinfo {pages} {49} (\bibinfo {year}
		{2022})}\BibitemShut {NoStop}%
	\bibitem [{\citenamefont {Paparini}\ and\ \citenamefont
		{Virga}(2022)}]{paparini:stability}%
	\BibitemOpen
	\bibfield  {author} {\bibinfo {author} {\bibfnamefont {S.}~\bibnamefont
			{Paparini}}\ and\ \bibinfo {author} {\bibfnamefont {E.~G.}\ \bibnamefont
			{Virga}},\ }\bibfield  {title} {\bibinfo {title} {Stability against the odds:
			the case of chromonic liquid crystals},\ }\href
	{https://doi.org/https://doi.org/10.1007/s00332-022-09833-6} {\bibfield
		{journal} {\bibinfo  {journal} {J. Nonlinear Sci.}\ }\textbf {\bibinfo
			{volume} {32}},\ \bibinfo {pages} {74} (\bibinfo {year} {2022})}\BibitemShut
	{NoStop}%
	\bibitem [{\citenamefont {Pedrini}\ and\ \citenamefont
		{Virga}(2020)}]{pedrini:liquid}%
	\BibitemOpen
	\bibfield  {author} {\bibinfo {author} {\bibfnamefont {A.}~\bibnamefont
			{Pedrini}}\ and\ \bibinfo {author} {\bibfnamefont {E.~G.}\ \bibnamefont
			{Virga}},\ }\bibfield  {title} {\bibinfo {title} {Liquid crystal distortions
			revealed by an octupolar tensor},\ }\href
	{https://doi.org/https://doi.org/10.1103/PhysRevE.101.012703} {\bibfield
		{journal} {\bibinfo  {journal} {Phys. Rev. E}\ }\textbf {\bibinfo {volume}
			{101}},\ \bibinfo {pages} {012703} (\bibinfo {year} {2020})}\BibitemShut
	{NoStop}%
	\bibitem [{\citenamefont {Gaeta}\ and\ \citenamefont
		{Virga}(2023)}]{gaeta:review}%
	\BibitemOpen
	\bibfield  {author} {\bibinfo {author} {\bibfnamefont {G.}~\bibnamefont
			{Gaeta}}\ and\ \bibinfo {author} {\bibfnamefont {E.~G.}\ \bibnamefont
			{Virga}},\ }\bibfield  {title} {\bibinfo {title} {A review on octupolar
			tensors},\ }\href {https://doi.org/10.1088/1751-8121/ace712} {\bibfield
		{journal} {\bibinfo  {journal} {J. Phys. A: Math. Theor.}\ }\textbf {\bibinfo
			{volume} {56}},\ \bibinfo {pages} {363001} (\bibinfo {year}
		{2023})}\BibitemShut {NoStop}%
	\bibitem [{\citenamefont {Meyer}(1976)}]{meyer:structural}%
	\BibitemOpen
	\bibfield  {author} {\bibinfo {author} {\bibfnamefont {R.~B.}\ \bibnamefont
			{Meyer}},\ }\bibfield  {title} {\bibinfo {title} {Structural problems in
			liquid crystal physics},\ }in\ \href@noop {} {\emph {\bibinfo {booktitle}
			{Molecular Fluids}}},\ \bibinfo {series} {Les {H}ouches {S}ummer {S}chool in
		{T}heoretical Physics}, Vol.\ \bibinfo {volume} {XXV-1973},\ \bibinfo
	{editor} {edited by\ \bibinfo {editor} {\bibfnamefont {R.}~\bibnamefont
			{Balian}}\ and\ \bibinfo {editor} {\bibfnamefont {G.}~\bibnamefont {Weill}}}\
	(\bibinfo  {publisher} {Gordon and Breach},\ \bibinfo {address} {New York},\
	\bibinfo {year} {1976})\ pp.\ \bibinfo {pages} {273--373}\BibitemShut
	{NoStop}%
	\bibitem [{\citenamefont {Cestari}\ \emph {et~al.}(2011)\citenamefont
		{Cestari}, \citenamefont {Diez-Berart}, \citenamefont {Dunmur}, \citenamefont
		{Ferrarini}, \citenamefont {de~la Fuente}, \citenamefont {Jackson},
		\citenamefont {Lopez}, \citenamefont {Luckhurst}, \citenamefont
		{Perez-Jubindo}, \citenamefont {Richardson}, \citenamefont {Salud},
		\citenamefont {Timimi},\ and\ \citenamefont {Zimmermann}}]{cestari:phase}%
	\BibitemOpen
	\bibfield  {author} {\bibinfo {author} {\bibfnamefont {M.}~\bibnamefont
			{Cestari}}, \bibinfo {author} {\bibfnamefont {S.}~\bibnamefont
			{Diez-Berart}}, \bibinfo {author} {\bibfnamefont {D.~A.}\ \bibnamefont
			{Dunmur}}, \bibinfo {author} {\bibfnamefont {A.}~\bibnamefont {Ferrarini}},
		\bibinfo {author} {\bibfnamefont {M.~R.}\ \bibnamefont {de~la Fuente}},
		\bibinfo {author} {\bibfnamefont {D.~J.~B.}\ \bibnamefont {Jackson}},
		\bibinfo {author} {\bibfnamefont {D.~O.}\ \bibnamefont {Lopez}}, \bibinfo
		{author} {\bibfnamefont {G.~R.}\ \bibnamefont {Luckhurst}}, \bibinfo {author}
		{\bibfnamefont {M.~A.}\ \bibnamefont {Perez-Jubindo}}, \bibinfo {author}
		{\bibfnamefont {R.~M.}\ \bibnamefont {Richardson}}, \bibinfo {author}
		{\bibfnamefont {J.}~\bibnamefont {Salud}}, \bibinfo {author} {\bibfnamefont
			{B.~A.}\ \bibnamefont {Timimi}},\ and\ \bibinfo {author} {\bibfnamefont
			{H.}~\bibnamefont {Zimmermann}},\ }\bibfield  {title} {\bibinfo {title}
		{Phase behavior and properties of the liquid-crystal dimer
			1$'${}$'${},7$'${}$'${}-bis(4-cyanobiphenyl-4$'${}-yl) heptane: A twist-bend
			nematic liquid crystal},\ }\href {https://doi.org/10.1103/PhysRevE.84.031704}
	{\bibfield  {journal} {\bibinfo  {journal} {Phys. Rev. E}\ }\textbf {\bibinfo
			{volume} {84}},\ \bibinfo {pages} {031704} (\bibinfo {year}
		{2011})}\BibitemShut {NoStop}%
	\bibitem [{\citenamefont {Borshch}\ \emph {et~al.}(2013)\citenamefont
		{Borshch}, \citenamefont {Kim}, \citenamefont {Xiang}, \citenamefont {Gao},
		\citenamefont {J\'{a}kli}, \citenamefont {Panov}, \citenamefont {Vij},
		\citenamefont {Imrie}, \citenamefont {Tamba}, \citenamefont {Mehl},\ and\
		\citenamefont {Lavrentovich}}]{borshch:nematic}%
	\BibitemOpen
	\bibfield  {author} {\bibinfo {author} {\bibfnamefont {V.}~\bibnamefont
			{Borshch}}, \bibinfo {author} {\bibfnamefont {Y.-K.}\ \bibnamefont {Kim}},
		\bibinfo {author} {\bibfnamefont {J.}~\bibnamefont {Xiang}}, \bibinfo
		{author} {\bibfnamefont {M.}~\bibnamefont {Gao}}, \bibinfo {author}
		{\bibfnamefont {A.}~\bibnamefont {J\'{a}kli}}, \bibinfo {author}
		{\bibfnamefont {V.~P.}\ \bibnamefont {Panov}}, \bibinfo {author}
		{\bibfnamefont {J.~K.}\ \bibnamefont {Vij}}, \bibinfo {author} {\bibfnamefont
			{C.~T.}\ \bibnamefont {Imrie}}, \bibinfo {author} {\bibfnamefont {M.~G.}\
			\bibnamefont {Tamba}}, \bibinfo {author} {\bibfnamefont {G.~H.}\ \bibnamefont
			{Mehl}},\ and\ \bibinfo {author} {\bibfnamefont {O.~D.}\ \bibnamefont
			{Lavrentovich}},\ }\bibfield  {title} {\bibinfo {title} {Nematic twist-bend
			phase with nanoscale modulation of molecular orientation},\ }\href
	{https://doi.org/https://doi.org/10.1038/ncomms3635} {\bibfield  {journal}
		{\bibinfo  {journal} {Nat. Commun.}\ }\textbf {\bibinfo {volume} {4}},\
		\bibinfo {pages} {2635} (\bibinfo {year} {2013})}\BibitemShut {NoStop}%
	\bibitem [{\citenamefont {Pollard}\ and\ \citenamefont
		{Alexander}(2021)}]{pollard:intrinsic}%
	\BibitemOpen
	\bibfield  {author} {\bibinfo {author} {\bibfnamefont {J.}~\bibnamefont
			{Pollard}}\ and\ \bibinfo {author} {\bibfnamefont {G.~P.}\ \bibnamefont
			{Alexander}},\ }\bibfield  {title} {\bibinfo {title} {Intrinsic geometry and
			director reconstruction for three-dimensional liquid crystals},\ }\href
	{https://doi.org/10.1088/1367-2630/abfdf4} {\bibfield  {journal} {\bibinfo
			{journal} {New J. Phys.}\ }\textbf {\bibinfo {volume} {23}},\ \bibinfo
		{pages} {063006} (\bibinfo {year} {2021})}\BibitemShut {NoStop}%
	\bibitem [{\citenamefont {{da~Silva}}\ and\ \citenamefont
		{Efrati}(2021)}]{dasilva:moving}%
	\BibitemOpen
	\bibfield  {author} {\bibinfo {author} {\bibfnamefont {L.~C.~B.}\
			\bibnamefont {{da~Silva}}}\ and\ \bibinfo {author} {\bibfnamefont
			{E.}~\bibnamefont {Efrati}},\ }\bibfield  {title} {\bibinfo {title} {Moving
			frames and compatibility conditions for three-dimensional director fields},\
	}\href {https://doi.org/10.1088/1367-2630/abfdf6} {\bibfield  {journal}
		{\bibinfo  {journal} {New J. Phys.}\ }\textbf {\bibinfo {volume} {23}},\
		\bibinfo {pages} {063016} (\bibinfo {year} {2021})}\BibitemShut {NoStop}%
	\bibitem [{\citenamefont {Niv}\ and\ \citenamefont
		{Efrati}(2018)}]{niv:geometric}%
	\BibitemOpen
	\bibfield  {author} {\bibinfo {author} {\bibfnamefont {I.}~\bibnamefont
			{Niv}}\ and\ \bibinfo {author} {\bibfnamefont {E.}~\bibnamefont {Efrati}},\
	}\bibfield  {title} {\bibinfo {title} {Geometric frustration and
			compatibility conditions for two-dimensional director fields},\ }\href
	{https://doi.org/https://doi.org/10.1039/C7SM01672G} {\bibfield  {journal}
		{\bibinfo  {journal} {Soft Matter}\ }\textbf {\bibinfo {volume} {14}},\
		\bibinfo {pages} {424} (\bibinfo {year} {2018})}\BibitemShut {NoStop}%
	\bibitem [{\citenamefont {Needham}(2021)}]{needham:visual}%
	\BibitemOpen
	\bibfield  {author} {\bibinfo {author} {\bibfnamefont {T.}~\bibnamefont
			{Needham}},\ }\href@noop {} {\emph {\bibinfo {title} {Visual Differential
				Geometry and Forms. {A} mathematical drama in five acts}}}\ (\bibinfo
	{publisher} {Princeton University Press},\ \bibinfo {address} {Princeton},\
	\bibinfo {year} {2021})\BibitemShut {NoStop}%
	\bibitem [{\citenamefont
		{Weatherburn}(2016{\natexlab{a}})}]{weatherburn:differential_1}%
	\BibitemOpen
	\bibfield  {author} {\bibinfo {author} {\bibfnamefont {C.~E.}\ \bibnamefont
			{Weatherburn}},\ }\href@noop {} {\emph {\bibinfo {title} {Differential
				Geometry of Three Dimensions}}},\ Vol.~\bibinfo {volume} {I}\ (\bibinfo
	{publisher} {Cambridge University Press},\ \bibinfo {address} {Cambridge},\
	\bibinfo {year} {2016})\BibitemShut {NoStop}%
	\bibitem [{\citenamefont
		{Weatherburn}(2016{\natexlab{b}})}]{weatherburn:differential_2}%
	\BibitemOpen
	\bibfield  {author} {\bibinfo {author} {\bibfnamefont {C.~E.}\ \bibnamefont
			{Weatherburn}},\ }\href@noop {} {\emph {\bibinfo {title} {Differential
				Geometry of Three Dimensions}}},\ Vol.~\bibinfo {volume} {II}\ (\bibinfo
	{publisher} {Cambridge University Press},\ \bibinfo {address} {Cambridge},\
	\bibinfo {year} {2016})\BibitemShut {NoStop}%
	\bibitem [{\citenamefont {Sonnet}\ and\ \citenamefont
		{Virga}(2024)}]{sonnet:bending-neutral}%
	\BibitemOpen
	\bibfield  {author} {\bibinfo {author} {\bibfnamefont {A.~M.}\ \bibnamefont
			{Sonnet}}\ and\ \bibinfo {author} {\bibfnamefont {E.~G.}\ \bibnamefont
			{Virga}},\ }\bibfield  {title} {\bibinfo {title} {Bending-neutral
			deformations of minimal surfaces},\ }\href
	{https://doi.org/10.1098/rspa.2024.0394} {\bibfield  {journal} {\bibinfo
			{journal} {Proc. Roy. Soc. Lond. A}\ }\textbf {\bibinfo {volume} {480}},\
		\bibinfo {pages} {20240394} (\bibinfo {year} {2024})}\BibitemShut {NoStop}%
	\bibitem [{\citenamefont {Burali-Forti}(1912)}]{burali-forti:fondamenti}%
	\BibitemOpen
	\bibfield  {author} {\bibinfo {author} {\bibfnamefont {C.}~\bibnamefont
			{Burali-Forti}},\ }\bibfield  {title} {\bibinfo {title} {Fondamenti per la
			geometria differenziale di una superficie col metodo vettoriale generale},\
	}\href {https://doi.org/https://doi.org/10.1007/BF03015286} {\bibfield
		{journal} {\bibinfo  {journal} {Rend. Circolo Mat. Palermo}\ }\textbf
		{\bibinfo {volume} {33}},\ \bibinfo {pages} {1} (\bibinfo {year}
		{1912})}\BibitemShut {NoStop}%
	\bibitem [{\citenamefont {Burgatti}(1917)}]{burgatti:teoremi}%
	\BibitemOpen
	\bibfield  {author} {\bibinfo {author} {\bibfnamefont {P.}~\bibnamefont
			{Burgatti}},\ }\bibfield  {title} {\bibinfo {title} {I teoremi del gradiente,
			della divergenza, della rotazione sopra una superficie e loro applicazione ai
			potenziali},\ }\href@noop {} {\bibfield  {journal} {\bibinfo  {journal}
			{Rend. Acc. Sci. Ist. Bologna}\ }\textbf {\bibinfo {volume} {4 (VII)}},\
		\bibinfo {pages} {3} (\bibinfo {year} {1917})}\BibitemShut {NoStop}%
	\bibitem [{\citenamefont {Burgatti}(1951)}]{burgatti:memorie}%
	\BibitemOpen
	\bibfield  {author} {\bibinfo {author} {\bibfnamefont {P.}~\bibnamefont
			{Burgatti}},\ }\bibinfo {title} {Memorie scelte}\ (\bibinfo  {publisher}
	{Zanichelli},\ \bibinfo {address} {Bologna},\ \bibinfo {year} {1951})\ pp.\
	\bibinfo {pages} {201--212}\BibitemShut {NoStop}%
	\bibitem [{\citenamefont {Burgatti}\ \emph {et~al.}(1930)\citenamefont
		{Burgatti}, \citenamefont {Boggio},\ and\ \citenamefont
		{Burali-Forti}}]{burgatti:analisi}%
	\BibitemOpen
	\bibfield  {author} {\bibinfo {author} {\bibfnamefont {P.}~\bibnamefont
			{Burgatti}}, \bibinfo {author} {\bibfnamefont {T.}~\bibnamefont {Boggio}},\
		and\ \bibinfo {author} {\bibfnamefont {C.}~\bibnamefont {Burali-Forti}},\
	}\href@noop {} {\emph {\bibinfo {title} {Analisi Vettoriale Generale. Vol.
				{II}: {G}eometria Differenziale}}}\ (\bibinfo  {publisher} {Zanichelli},\
	\bibinfo {address} {Bologna},\ \bibinfo {year} {1930})\BibitemShut {NoStop}%
	\bibitem [{\citenamefont {Rosso}\ \emph {et~al.}(2012)\citenamefont {Rosso},
		\citenamefont {Virga},\ and\ \citenamefont {Kralj}}]{rosso:parallel}%
	\BibitemOpen
	\bibfield  {author} {\bibinfo {author} {\bibfnamefont {R.}~\bibnamefont
			{Rosso}}, \bibinfo {author} {\bibfnamefont {E.~G.}\ \bibnamefont {Virga}},\
		and\ \bibinfo {author} {\bibfnamefont {S.}~\bibnamefont {Kralj}},\ }\bibfield
	{title} {\bibinfo {title} {Parallel transport and defects on nematic
			shells},\ }\href {https://doi.org/https://doi.org/10.1007/s00161-012-0259-4}
	{\bibfield  {journal} {\bibinfo  {journal} {Continuum Mech. Thermodyn.}\
		}\textbf {\bibinfo {volume} {24}},\ \bibinfo {pages} {643} (\bibinfo {year}
		{2012})}\BibitemShut {NoStop}%
	\bibitem [{\citenamefont {Truesdell}(1991)}]{truesdell:first}%
	\BibitemOpen
	\bibfield  {author} {\bibinfo {author} {\bibfnamefont {C.~A.}\ \bibnamefont
			{Truesdell}},\ }\href@noop {} {\emph {\bibinfo {title} {A First Course in
				Rational Continuum Mechanics}}},\ \bibinfo {edition} {2nd}\ ed.,\ \bibinfo
	{series} {Pure and Applied Mathematics}, Vol.~\bibinfo {volume} {71}\
	(\bibinfo  {publisher} {Academic Press},\ \bibinfo {address} {Boston},\
	\bibinfo {year} {1991})\BibitemShut {NoStop}%
	\bibitem [{\citenamefont {Cartan}(1935)}]{cartan:methode}%
	\BibitemOpen
	\bibfield  {author} {\bibinfo {author} {\bibfnamefont {E.}~\bibnamefont
			{Cartan}},\ }\href@noop {} {\emph {\bibinfo {title} {La M\'ethode du Rep\`ere
				Mobile, La Th\'eorie des Groupes Continue et Les Espaces
				G\'en\'eralis\'es}}},\ \bibinfo {series} {Actualit\'es Scientifiques et
		Industrielles}, Vol.\ \bibinfo {volume} {194}\ (\bibinfo  {publisher}
	{Hermann},\ \bibinfo {address} {Paris},\ \bibinfo {year} {1935})\ \bibinfo
	{note} {available from
		\url{{https://gallica.bnf.fr/ark:/12148/bpt6k38182z}}}\BibitemShut {NoStop}%
	\bibitem [{\citenamefont {O'{N}eill}(2006)}]{o'neill:elementary}%
	\BibitemOpen
	\bibfield  {author} {\bibinfo {author} {\bibfnamefont {B.}~\bibnamefont
			{O'{N}eill}},\ }\href@noop {} {\emph {\bibinfo {title} {Elementary
				Differential Geometry}}},\ \bibinfo {edition} {2nd}\ ed.\ (\bibinfo
	{publisher} {Academic Press},\ \bibinfo {address} {Burlington},\ \bibinfo
	{year} {2006})\BibitemShut {NoStop}%
	\bibitem [{\citenamefont {Clelland}(2017)}]{clelland:from}%
	\BibitemOpen
	\bibfield  {author} {\bibinfo {author} {\bibfnamefont {J.~N.}\ \bibnamefont
			{Clelland}},\ }\href@noop {} {\emph {\bibinfo {title} {From {F}renet to
				{C}artan: {T}he Method of Moving Frames}}},\ \bibinfo {series} {Graduate
		Studies in Mathematics}, Vol.\ \bibinfo {volume} {178}\ (\bibinfo
	{publisher} {American Mathematical Society},\ \bibinfo {address}
	{Providence},\ \bibinfo {year} {2017})\BibitemShut {NoStop}%
	\bibitem [{\citenamefont {Ozenda}\ \emph {et~al.}(2020)\citenamefont {Ozenda},
		\citenamefont {Sonnet},\ and\ \citenamefont {Virga}}]{ozenda:blend}%
	\BibitemOpen
	\bibfield  {author} {\bibinfo {author} {\bibfnamefont {O.}~\bibnamefont
			{Ozenda}}, \bibinfo {author} {\bibfnamefont {A.~M.}\ \bibnamefont {Sonnet}},\
		and\ \bibinfo {author} {\bibfnamefont {E.~G.}\ \bibnamefont {Virga}},\
	}\bibfield  {title} {\bibinfo {title} {A blend of stretching and bending in
			nematic polymer networks},\ }\href
	{https://doi.org/https://doi.org/10.1039/D0SM00642D} {\bibfield  {journal}
		{\bibinfo  {journal} {Soft Matter}\ }\textbf {\bibinfo {volume} {16}},\
		\bibinfo {pages} {8877} (\bibinfo {year} {2020})}\BibitemShut {NoStop}%
	\bibitem [{\citenamefont {Ozenda}\ and\ \citenamefont
		{Virga}(2021)}]{ozenda:kirchhoff}%
	\BibitemOpen
	\bibfield  {author} {\bibinfo {author} {\bibfnamefont {O.}~\bibnamefont
			{Ozenda}}\ and\ \bibinfo {author} {\bibfnamefont {E.~G.}\ \bibnamefont
			{Virga}},\ }\bibfield  {title} {\bibinfo {title} {On the {K}irchhoff-{L}ove
			hypothesis (revised and vindicated)},\ }\href
	{https://doi.org/https://doi.org/10.1007/s10659-021-09819-7} {\bibfield
		{journal} {\bibinfo  {journal} {J. Elast.}\ }\textbf {\bibinfo {volume}
			{143}},\ \bibinfo {pages} {359} (\bibinfo {year} {2021})}\BibitemShut
	{NoStop}%
	\bibitem [{\citenamefont {Sonnet}\ and\ \citenamefont
		{Virga}(2025)}]{sonnet:variational}%
	\BibitemOpen
	\bibfield  {author} {\bibinfo {author} {\bibfnamefont {A.~M.}\ \bibnamefont
			{Sonnet}}\ and\ \bibinfo {author} {\bibfnamefont {E.~G.}\ \bibnamefont
			{Virga}},\ }\bibfield  {title} {\bibinfo {title} {A variational theory for
			soft shells},\ }\href
	{https://doi.org/https://doi.org/10.1016/j.jmps.2025.106132} {\bibfield
		{journal} {\bibinfo  {journal} {J. Mech. Phys. Solids}\ }\textbf {\bibinfo
			{volume} {200}},\ \bibinfo {pages} {106132} (\bibinfo {year}
		{2025})}\BibitemShut {NoStop}%
	\bibitem [{\citenamefont {do~{C}armo}(2016)}]{doCarmo:differential}%
	\BibitemOpen
	\bibfield  {author} {\bibinfo {author} {\bibfnamefont {M.~P.}\ \bibnamefont
			{do~{C}armo}},\ }\href@noop {} {\emph {\bibinfo {title} {Differential
				Geometry of Curves and Surfaces}}},\ \bibinfo {edition} {2nd}\ ed.\ (\bibinfo
	{publisher} {Dover},\ \bibinfo {address} {Mineola},\ \bibinfo {year}
	{2016})\BibitemShut {NoStop}%
	\bibitem [{\citenamefont {Levi-Civita}(1916)}]{levi-civita:nozione}%
	\BibitemOpen
	\bibfield  {author} {\bibinfo {author} {\bibfnamefont {T.}~\bibnamefont
			{Levi-Civita}},\ }\bibfield  {title} {\bibinfo {title} {Nozione di
			parallelismo in una variet\`a qualunque e conseguente specificazione
			geometrica della curvatura riemanniana},\ }\href
	{https://doi.org/https://doi.org/10.1007/BF03014898} {\bibfield  {journal}
		{\bibinfo  {journal} {Rend. Circolo Mat. Palermo}\ }\textbf {\bibinfo
			{volume} {42}},\ \bibinfo {pages} {173} (\bibinfo {year} {1916})}\BibitemShut
	{NoStop}%
	\bibitem [{\citenamefont {Persico}(1921)}]{persico:realizzazione}%
	\BibitemOpen
	\bibfield  {author} {\bibinfo {author} {\bibfnamefont {E.}~\bibnamefont
			{Persico}},\ }\bibfield  {title} {\bibinfo {title} {Realizzazione cinematica
			del parallelismo superficiale},\ }\href@noop {} {\bibfield  {journal}
		{\bibinfo  {journal} {Atti R. Acc. Linc. Rend. Cl. Scienze Mat. Fis. Nat.}\
		}\textbf {\bibinfo {volume} {30 (V)}},\ \bibinfo {pages} {127} (\bibinfo
		{year} {1921})},\ \bibinfo {note} {available from
		\url{http://operedigitali.lincei.it/rendicontiFMN/rol/visabs.php?lang=it&type=mat&fileId=5947}}\BibitemShut
	{NoStop}%
	\bibitem [{\citenamefont {Chern}(1955)}]{chern:elementary}%
	\BibitemOpen
	\bibfield  {author} {\bibinfo {author} {\bibfnamefont {S.-S.}\ \bibnamefont
			{Chern}},\ }\bibfield  {title} {\bibinfo {title} {An elementary proof of the
			existence of isothermal parameters on a surface},\ }\href
	{https://doi.org/https://doi.org/10.1090/S0002-9939-1955-0074856-1}
	{\bibfield  {journal} {\bibinfo  {journal} {Proc. Amer. Math. Soc.}\ }\textbf
		{\bibinfo {volume} {6}},\ \bibinfo {pages} {771} (\bibinfo {year}
		{1955})}\BibitemShut {NoStop}%
	\bibitem [{\citenamefont {Man}\ and\ \citenamefont
		{Cohen}(1986)}]{man:coordinate}%
	\BibitemOpen
	\bibfield  {author} {\bibinfo {author} {\bibfnamefont {C.-S.}\ \bibnamefont
			{Man}}\ and\ \bibinfo {author} {\bibfnamefont {H.}~\bibnamefont {Cohen}},\
	}\bibfield  {title} {\bibinfo {title} {A coordinate-free approach to the
			kinematics of membranes},\ }\href
	{https://doi.org/https://doi.org/10.1007/BF00041068} {\bibfield  {journal}
		{\bibinfo  {journal} {J. Elast.}\ }\textbf {\bibinfo {volume} {16}},\
		\bibinfo {pages} {97} (\bibinfo {year} {1986})}\BibitemShut {NoStop}%
	\bibitem [{\citenamefont {Pressley}(2012)}]{pressley:elementary}%
	\BibitemOpen
	\bibfield  {author} {\bibinfo {author} {\bibfnamefont {A.}~\bibnamefont
			{Pressley}},\ }\href@noop {} {\emph {\bibinfo {title} {Elementary
				Differential Geometry}}},\ \bibinfo {edition} {2nd}\ ed.,\ Springer
	Undergraduate Mathematics Series\ (\bibinfo  {publisher} {Springer-Verlag},\
	\bibinfo {address} {London},\ \bibinfo {year} {2012})\BibitemShut {NoStop}%
	\bibitem [{\citenamefont {Dorfmeister}\ and\ \citenamefont
		{Sterling}(2014)}]{dorfmeister:minding}%
	\BibitemOpen
	\bibfield  {author} {\bibinfo {author} {\bibfnamefont {J.~F.}\ \bibnamefont
			{Dorfmeister}}\ and\ \bibinfo {author} {\bibfnamefont {I.}~\bibnamefont
			{Sterling}},\ }\bibfield  {title} {\bibinfo {title} {Minding’s theorem for
			low degrees of differentiability},\ }\href
	{https://doi.org/10.3836/TJM/1422452805} {\bibfield  {journal} {\bibinfo
			{journal} {Tokyo J. Math.}\ }\textbf {\bibinfo {volume} {37}},\ \bibinfo
		{pages} {503} (\bibinfo {year} {2014})}\BibitemShut {NoStop}%
	\bibitem [{\citenamefont {B\"ohmer}\ \emph {et~al.}(2020)\citenamefont
		{B\"ohmer}, \citenamefont {Lee},\ and\ \citenamefont
		{Neff}}]{bohmer:chirality}%
	\BibitemOpen
	\bibfield  {author} {\bibinfo {author} {\bibfnamefont {C.~G.}\ \bibnamefont
			{B\"ohmer}}, \bibinfo {author} {\bibfnamefont {Y.}~\bibnamefont {Lee}},\ and\
		\bibinfo {author} {\bibfnamefont {P.}~\bibnamefont {Neff}},\ }\bibfield
	{title} {\bibinfo {title} {Chirality in the plane},\ }\href
	{https://doi.org/https://doi.org/10.1016/j.jmps.2019.103753} {\bibfield
		{journal} {\bibinfo  {journal} {J. Mech. Phys. Solids}\ }\textbf {\bibinfo
			{volume} {134}},\ \bibinfo {pages} {103753} (\bibinfo {year}
		{2020})}\BibitemShut {NoStop}%
\end{thebibliography}

%

\end{document}